\documentclass[%
reprint,
nofootinbib,
nobibnotes,
aps,
prl,
]{revtex4-2}

\usepackage{graphicx}
\usepackage{dcolumn}
\usepackage{bm}
\usepackage{hyperref}
\usepackage{multirow}

\begin{document}

\title{Improved limits for violations of local position invariance from atomic clock comparisons}
\author{R.~Lange}%
\author{N.~Huntemann}%
\email{nils.huntemann@ptb.de}
\author{J.~M.~Rahm}%
\author{C.~Sanner}
\altaffiliation[Present address: ]{JILA, Boulder, CO 80309, USA} 
\author{H.~Shao}
\author{B.~Lipphardt}
\author{Chr.~Tamm}
\author{S.~Weyers}
\author{E.~Peik}
\affiliation{%
 Physikalisch-Technische Bundesanstalt, Bundesallee 100, 38116 Braunschweig, Germany
}%

\date{\today}

\begin{abstract}
We compare two optical clocks based on the $^2$S$_{1/2}(F=0)\to {}^2$D$_{3/2}(F=2)$ electric quadrupole (E2) and the $^2$S$_{1/2}(F=0)\to {}^2$F$_{7/2}(F=3)$ electric octupole (E3) transition of \textsuperscript{171}Yb\textsuperscript{+} and measure the frequency ratio $\nu_{\mathrm{E3}}/\nu_{\mathrm{E2}}=0.932\,829\,404\,530\,965\,376(32)$, improving upon previous measurements by an order of magnitude. Using two caesium fountain clocks, we find $\nu_{E3}=642\,121\,496\,772\,645.10(8)$~Hz, the most accurate determination of an optical transition frequency to date. Repeated measurements of both quantities over several years are analyzed for potential violations of local position invariance. We improve by factors of about 20 and 2 the limits for fractional temporal variations of the fine structure constant $\alpha$ to $1.0(1.1)\times10^{-18}/\mathrm{yr}$ and of the proton-to-electron mass ratio $\mu$ to $-8(36)\times10^{-18}/\mathrm{yr}$. Using the annual variation of the Sun's gravitational potential at Earth $\Phi$, we improve limits for a potential coupling of both constants to gravity, $(c^2/\alpha) (d\alpha/d\Phi)=14(11)\times 10^{-9}$ and $(c^2/\mu) (d\mu/d\Phi)=7(45)\times 10^{-8}$.
\end{abstract}

\maketitle
Searches for violations of Einstein's equivalence principle, such as tests of local Lorentz invariance (LLI) and local position invariance (LPI), have become one of the leading applications of low-energy high-precision experiments with laser-cooled atoms or ions \cite{saf18a}. As part of the Einstein equivalence principle, LPI states that the result of any non-gravitational experiment is independent of the position in space and time \cite{wil14}. 
While theories beyond the standard model predict temporal  variations of fundamental constants  \cite{mar17} and astronomical observations indicate a spatial variation of the fine structure constant $\alpha$ \cite{wil20}, no experimental observation of any violation of LPI in a laboratory setting has been reported so far \cite{saf18a}. 
Promising test cases in tabletop experiments are comparisons of atomic clocks based on transitions that show a different dependence of their frequency on the value of fundamental constants \cite{ros08,god14,hun14,gue12,ash18}. 
Here, a potential variation would become observable as a change in the frequency ratio of the clocks, that nowadays reach fractional frequency uncertainties of one part in $10^{18}$ and below \cite{mcg18, bot19, san19, bre19}. 
While a small frequency uncertainty of the clock is a prerequisite to reveal undetected indications of physics beyond the standard model, it is of similar importance to have a large sensitivity of the measured quantity, i.e. the frequency ratio of the clocks used in the search, to the potentially varying constant. 
Particularly large magnification of relative shifts of $\alpha$ is expected from reference transitions in highly-charged ions \cite{koz18} and even larger for the low-energy nuclear transition in $^{229}$Th \cite{fla06,thi18}. 
The so far most stringent experimental limit on a temporal drift of $\alpha$ is based on the frequency ratio of Al$^+$ and Hg$^+$ single-ion optical clocks, repeatedly measured over a period of one year in 2006-2007 \cite{ros08}.

In this Letter, we report a more stringent test of LPI and investigate variations of fundamental constants using comparisons between optical atomic clocks based on ytterbium ions and microwave caesium fountain clocks over a period of more than 4 years. 
The optical clocks employ the $^{2}$S$_{1/2} (F=0)\to {}^2$D$_{3/2}(F=2)$ electric quadrupole (E2) or the $^{2}$S$_{1/2} (F=0)\to {}^2$F$_{7/2}(F=3)$ electric octupole (E3) transition of a single trapped \textsuperscript{171}Yb$^+$ ion. 
For the E2 transition, the natural lifetime of the excited state of about 50~ms limits the interrogation time and we typically use a single 40~ms Rabi pulse. 
The excited state lifetime of several years on the E3 transition \cite{rob00} enables us to take full advantage of the available laser coherence time. 
With means of an optical frequency comb, we transfer the excellent frequency instability of a laser stabilized to a single-crystal cryogenic silicon cavity \cite{mat17a} to the probe laser systems and realize coherent interrogation times of up to 500~ms.

As already indicated by the different natural lifetimes, the electronic structures of the excited states of the E2 and E3 transition differ significantly.
They are characterized by a single $6d$ electron for the ${}^2$D$_{3/2}$ state and a single hole in the otherwise filled $4f$ shell for the ${}^2$F$_{7/2}$ state. 
The proximity of the $4f$ shell to the nucleus of this heavy ion makes it plausible that the E3 transition energy possesses large relativistic contributions, making this transition frequency the most sensitive to changes of $\alpha$ among the presently operational optical clocks \cite{fla09}. 
Furthermore, the opportunity to compare two different transition frequencies provided by the same ion reduces the complexity of the experimental apparatus.
 
The two independent clock systems used for our investigation have recently been evaluated to below $3\times 10^{-18}$ fractional uncertainty on the E3 transition. In a comparison over a period of six months, they showed an agreement within their combined uncertainty and provided a stringent test of LLI \cite{san19}. 
Here we operate one clock system on the E3 and the other on the E2 transition to measure the frequency ratio $\nu_{\text{E3}}/\nu_{\text{E2}}$ and to perform tests of LPI. 

We employ rotationally symmetric radio frequency Paul traps to confine single $^{171}$Yb$^+$ ions, laser-cooled on the $^{2}$S$_{1/2}\to {}^2$P$_{1/2}$ transition at 370~nm to the Doppler temperature limit at below 1~mK. 
Population trapping due to spontaneous decay, background gas collisions or successful excitation of the clock transitions is prevented by state repumping with 935~nm and 760~nm laser light.
A large magnetic field of about 0.5~mT prevents coherent population trapping during the cooling period \cite{ejt10}, while a well-controlled magnetic field of a few $\mu$T during interrogation of the clock transition defines the orientation of the quantization axis \cite{lan20}. 
Averaging over three mutually orthogonal orientations of the small magnetic field, tensorial shifts are measured and suppressed on the E2 transition \cite{ita00}. 
For the E3 transition, the corresponding shift along a given orientation is inferred from the measured E2 shift and the known relative sensitivity of both transitions \cite{lan20}.
Frequency shifts due to residual ac magnetic fields from the radio frequency trap field as reported in \cite{gan18} have been investigated and an upper bound is given in the uncertainty budget shown in table~\ref{tab:E2uncert}. 
There, the frequency shifts $\delta\nu_{\mathrm{E2}}$ and corresponding uncertainties evaluated for the E2 transition are summarized, yielding a total uncertainty of $3.3\times10^{-17}$. 
The leading uncertainty contribution caused by the frequency shift induced by thermal radiation has been substantially reduced using an investigation of the differential polarizability performed at the National Physical Laboratory \cite{bay18}, resulting in the significant improvement of the total uncertainty compared to our previously published value of $1.1\times10^{-16}$ \cite{tam14}.

Since most operational parameters have been chosen equally over the measurement period, the reproducibility of frequency shifts reported in table~\ref{tab:E2uncert} is expected to be significantly better than the corresponding systematic uncertainty. 
Furthermore, uncertainties in atomic parameters, such as the differential polarizability, do not contribute to the reproducibility directly. 
For the E3 transition, frequency shifts $\delta\nu_{\mathrm{E3}}/\nu_{\mathrm{E3}}$ have been evaluated to below $3\times10^{-18}$ total uncertainty \cite{san19} and contribute negligibly to the uncertainty of the comparison of the two clocks.  

\begin{table}[ht]
    \centering
    \begin{ruledtabular}
    \begin{tabular}{l|c|c|c}
    Effect  & {$\delta\nu_{\mathrm{E2}} / \nu_{\mathrm{E2}}$}& {$u_{\mathrm{E2}} / \nu_{\mathrm{E2}}$}&{$r_{\mathrm{E2}} / \nu_{\mathrm{E2}}$} \\
            & {$\left(10^{-18}\right) $}&{$\left(10^{-18}\right) $}&{$\left(10^{-18}\right)$}\\ \hline
    Blackbody radiation         & -495                          & 27                        & 3\\
    Quadrupole                  & 0                             & 14                        & 14\\
    Second-order Zeeman (rf)    & 0                             & 10                        & 1\\
    Second-order Zeeman (dc)    & 463                           & 6                         & 6\\
    Second-order Stark          & -2.0                          & 2                         & 2\\
    Second-order Doppler        & -1.0                          & 1                         & 1\\
    Servo                       & 0                             & 1                         & 1\\ \hline
    Total                       & -35                           & 33                        & 16\\ 
    \end{tabular}
    \end{ruledtabular}
    \caption{Leading frequency shift effects evaluated for the single ion clock using the $^{2}$S$_{1/2} (F=0)\to {}^2$D$_{3/2}(F=2)$ electric quadrupole (E2) transition of $^{171}$Yb$^+$. The frequency shifts $\delta\nu_{\mathrm{E2}}$, the corresponding uncertainty $u_{E2}$, and the reproducibility $r_{\mathrm{E2}}$ are given in fractional units of the unperturbed transition frequency $\nu_{\mathrm{E2}}$. The second-order Zeeman effect is separated into shifts resulting from radio frequency (rf) and quasi-constant (dc) fields. }
    \label{tab:E2uncert}
\end{table}

\begin{figure}
{\centering \includegraphics[width=.95\columnwidth]{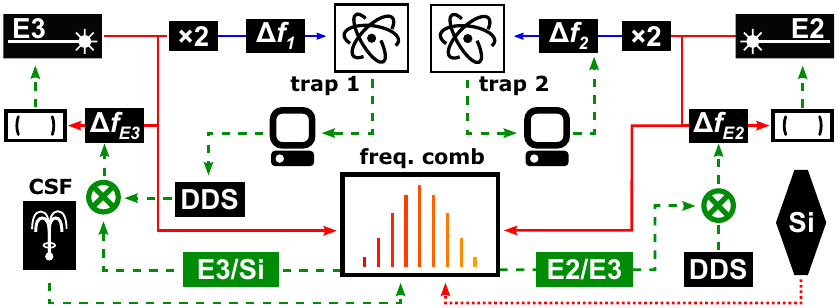}}
\caption{Schematic setup for the clock comparisons with optical paths depicted by solid red and blue lines and electric signals shown by dashed green lines. Both the E3 and E2 probe lasers are locked with a bandwidth of about 500~kHz to optical cavities. The laser light is sent to a frequency comb and, after frequency doubling, to the ion traps. Frequency offsets $\Delta f_1$ and $\Delta f_2$ are applied during clock operation. The comb also receives signals from the caesium fountains (CSF) and, indicated by the dotted red line, from a laser locked to an ultra-stable silicon cavity (Si) \cite{mat17a}. The Si cavity enhances the short-term stability of the E3 laser. The E2 laser frequency is controlled to provide a constant frequency ratio $1/R_0$ relative to the E3 laser at the frequency comb.
Spectroscopy of the E3 transition of the ion in trap 1 steers the E3 laser frequency via direct digital synthesis (DDS), while the frequency of the E2 laser applied to trap 2 is corrected by an additional offset to $\Delta f_2$.}
\label{graph:Setup}
\end{figure}

The ratio of the two reference transitions is measured by means of an optical frequency comb generator as shown in Fig.~\ref{graph:Setup}. 
To improve the short-term frequency instability of the E3 probe laser system, it is stabilized using the transfer oscillator concept \cite{ste02a} to a laser system referencing the cryogenic silicon cavity Si-2 \cite{mat17a}. 
Using the discriminator signal obtained by spectroscopy of the E3 transition, the ratio of the E3 clock laser frequency and that of the laser stabilized to the Si-2 cavity is adjusted using a digital second-order integrating servo system. 
Small frequency offsets required to compensate for the ac Stark shift of the probe laser in the interrogation sequence and to generate the discriminator signal are applied to an acousto-optic modulator (AOM) in front of the ion trap setup \cite{hun16}.
In this way, the probe laser light sent to the frequency comb is, up to a constant frequency offset and the correction for systematic frequency shifts, at the clock output frequency which approximates the unperturbed transition frequency. 
In contrast to the E3 spectroscopy, the E2 probe laser frequency at the frequency comb is stabilized with fixed ratio $1/R_0$ to that of the E3 probe laser. 
Based on spectroscopy of the E2 transition, the frequency of the AOM in front of the ion is corrected using a digital integrating servo system. 
The applied correction $\Delta\nu_{\mathrm{E2}}$ is recorded and permits calculation of 

\begin{eqnarray}
        R_{\mathrm{E3,E2}} &=& \nu_{\mathrm{E3}}/\nu_{\mathrm{E2}} \nonumber \\
     &=& R_0 \left(1+\frac{\Delta\nu_{\mathrm{E2}}}{\nu_{\mathrm{E2}}}+\frac{\delta\nu_{\mathrm{E2}}}{\nu_{\mathrm{E2}}}-\frac{\delta\nu_{\mathrm{E3}}}{\nu_{\mathrm{E3}}} \right).
\end{eqnarray}

\begin{figure}
{\centering \includegraphics[width=.95\columnwidth]{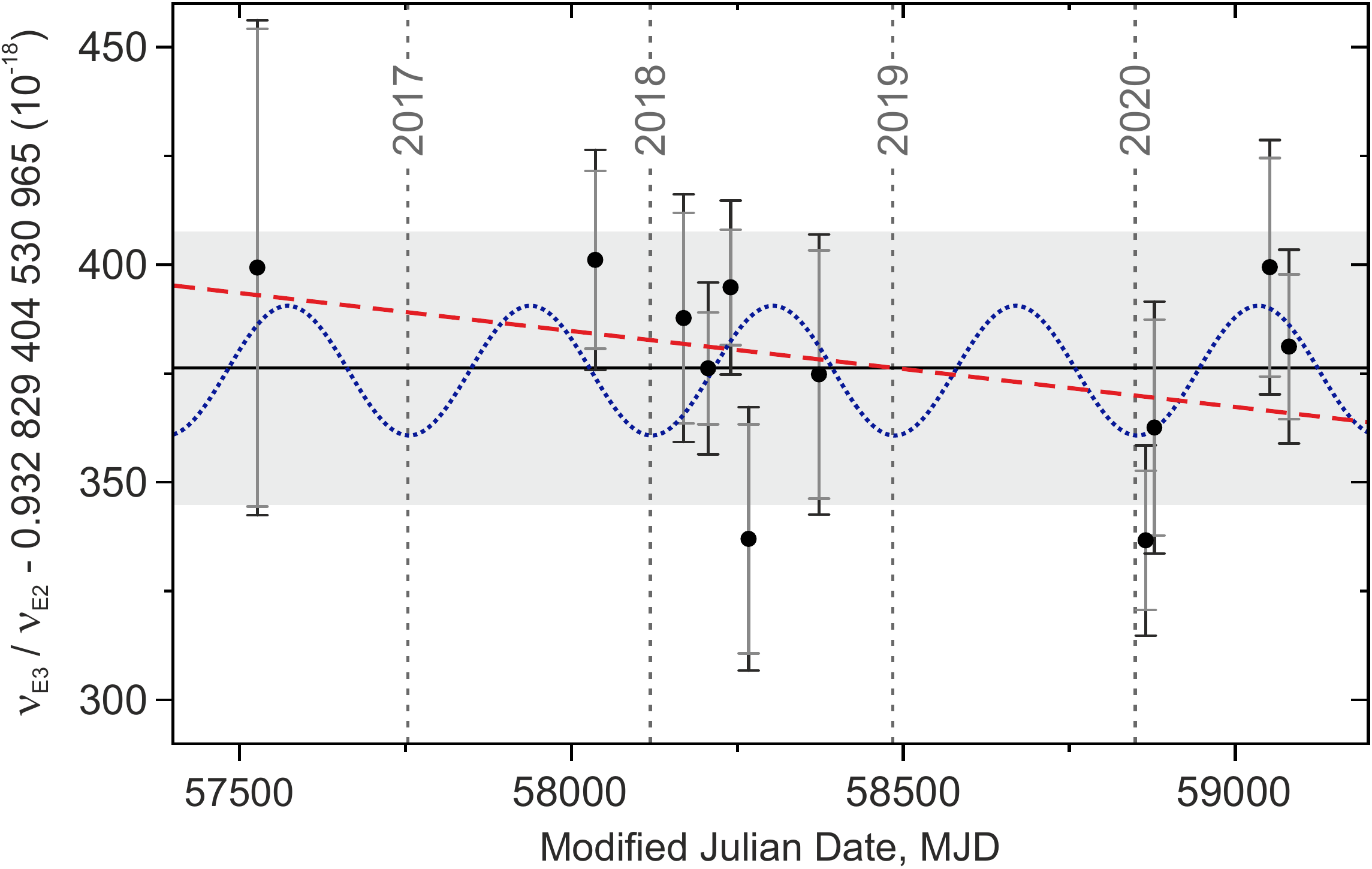}}
\caption{Ratio of the frequencies $\nu_{\mathrm{E3}}$ and $\nu_{\mathrm{E2}}$ of the electric octupole and the electric quadrupole transition measured over a period of $1\,500$ days between MJD 57527 (May 19, 2016) and MJD 59081 (August 20, 2020). The inner grey error bars show the statistical uncertainties. For the outer black error bars, a $1.6\times10^{-17}$ fractional uncertainty has been added in quadrature to take into account the reproducibility of systematic shifts over the measurement period (see text). The solid line gives the weighted average and the grey shaded area shows the total fractional uncertainty of $34\times 10^{-18}$. The dashed red line and the dotted blue line are fits to the data for searches for a temporal drift and a dependence on the gravitational potential.} 
\label{graph:RatiovsTime}
\end{figure}

The results of 11 measurements of $R_{\mathrm{E3,E2}}$ performed over a period of $1\,500$ days are shown in  Fig.~\ref{graph:RatiovsTime}.
Taking into account only the statistical uncertainty of each measurement yields a reduced chi-square value of $\chi^2_\mathrm{red} = 1.3$.
Including the fractional reproducibility of the E2 transition of $1.6\times 10^{-17}$ yields $\chi^2_\mathrm{red} = 0.8$.
The weighted average of the complete data set gives $R_{\mathrm{E3,E2}} = 0.932\,829\,404\,530\,965\,376(32)$. The total fractional uncertainty of $34\times 10^{-18}$ is dominated by the systematic uncertainty of the E2 transition $(33\times 10^{-18})$. The statistical contribution $(6\times 10^{-18})$ and the systematic uncertainty of the E3 transition $(3\times 10^{-18})$ are negligible. This result differs by $2.3$ standard uncertainties from a previous measurement and improves the uncertainty by one order of magnitude \cite{god14}. 

\begin{figure}
{\centering \includegraphics[width=.95\columnwidth]{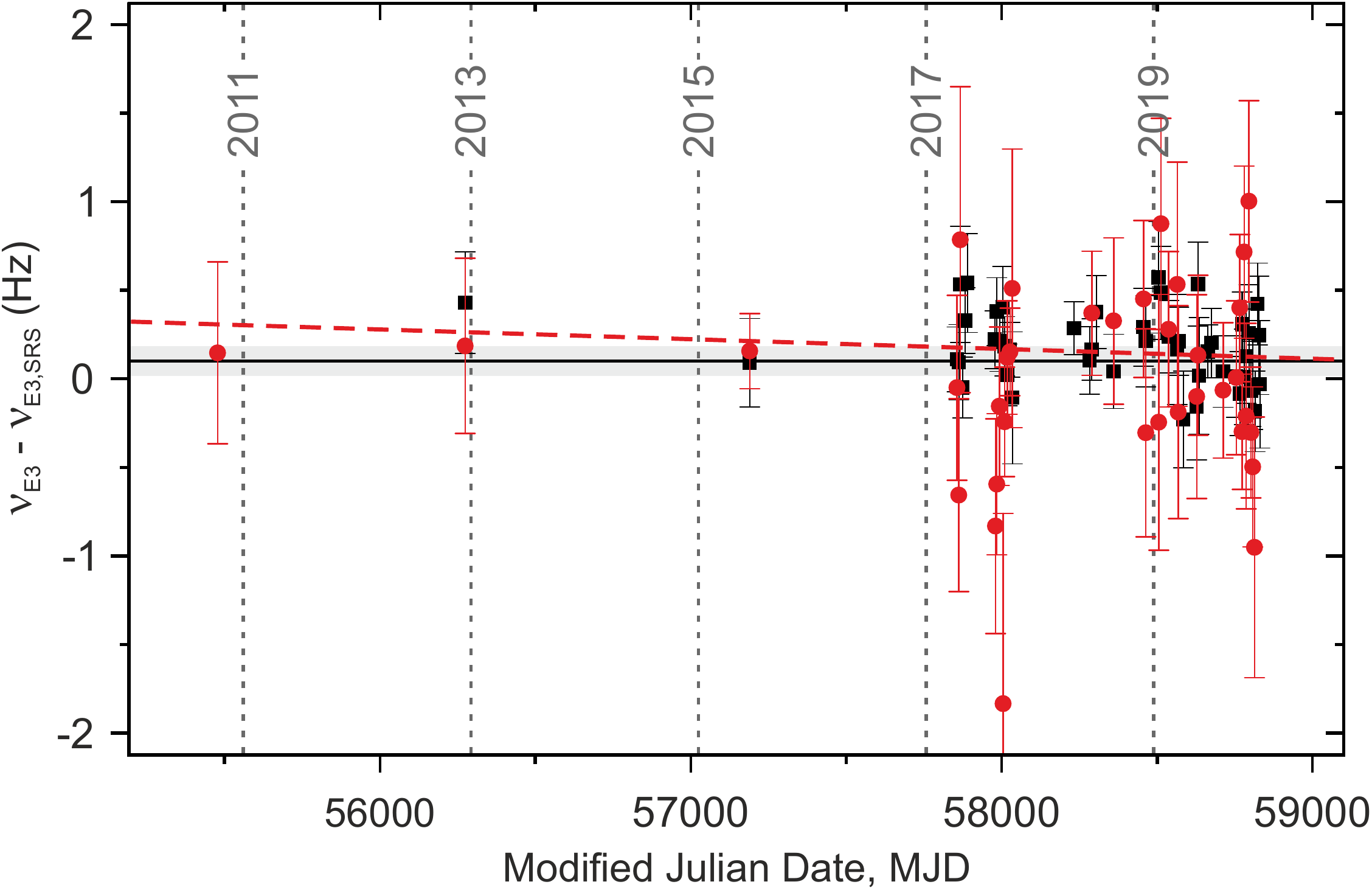}}
\caption{Measurements of the frequency of the electric octupole transition $\nu_{\mathrm{E3}}$ conducted between MJD 55479 (October 10, 2010) and MJD 58830 (December 13, 2019). Data is shown relative to the recommended value of the transition frequency $\nu_{\mathrm{E3, SRS}}=642\,121\,496\,772\,645$~Hz \cite{rie18}. Red dots and black squares are results obtained with caesium fountain clocks CSF1 and CSF2, respectively, that serve as primary frequency standards of our laboratory. The error bars indicate $1\sigma$ total uncertainties. The measurements centered at MJD 55479 and MJD 56275 are published in Refs.~\cite{hun12,hun14}. The solid line shows the weighted average of the data since MJD 57854 and the grey shaded area shows the corresponding total fractional uncertainty of $1.3\times 10^{-16}$. The red dashed line is a linear fit to the full data set for searches for a temporal drift.} 
\label{graph:FreqvsTime}
\end{figure}

Besides the measurement of the optical frequency ratio, the E3 transition frequency is measured with the microwave caesium fountain clocks CSF1 and CSF2 of our laboratory \cite{wey18}. 
To increase the total measurement time, interruptions shorter than one day in the operation of the optical clock are bridged by using a hydrogen maser as a flywheel oscillator \cite{leu16,gre16}. 
Numerical simulations of the typically observed frequency instability of the H-maser allow us to evaluate the corresponding uncertainty contribution.
Figure~\ref{graph:FreqvsTime} shows this data together with results previously acquired using the same clocks: The measurements centered at Modified Julian Date (MJD) 55479 and MJD 56275 have been published \cite{hun12,hun14} and the values at MJD 57190 were obtained during a European clock comparison campaign \cite{rie20}. 
The frequency of the E3 transition is determined as the weighted average of the contribution of each caesium fountain clock. 
The weights $w_i$ of each data point correspond to the inverse squared-sum $1/(u_{a,i}^2+u_{b,i}^2)$ of the systematic uncertainty $u_{b,i}$ and the statistical uncertainty $u_{a,i}$ that includes a contribution from the extrapolation using the H-maser. 
From data recorded between MJD 57854 and MJD 58830, we find an averaged value $\nu_{\mathrm{E3}}=642\,121\,496\,772\,645.10(8)$~Hz obtained from individual results with CSF1 and CSF2 of $\nu_{\mathrm{E3}}-0.09(13)$~Hz and $\nu_{\mathrm{E3}}+0.08(11)$~Hz with $2\,478$ and $4\,394$ hours of measurement. 
The CSF1 data yields $\chi^2_\mathrm{red} = 1.0$ and the CSF2 data $\chi^2_\mathrm{red} = 0.9$. 
The total fractional uncertainty of $1.3\times 10^{-16}$ is dominated by the systematic uncertainties of $1.6\times10^{-16}$ and $1.7\times10^{-16}$ of CSF1 and CSF2 \cite{wey18,sch20d}. 
The result is in excellent agreement with previous measurements and reduces the uncertainty in the frequency by more than a factor of 3 \cite{hun14,bay18a}.
To our knowledge $\nu_{\mathrm{E3}}$ is the third optical transition for which the absolute frequency is reported with a fractional uncertainty below $2\times10^{-16}$ \cite{mcg19,sch20d,nem20} and our result represents the most accurate measurement of an optical transition frequency to date. With the frequency ratio determined above, the measurement of $\nu_{\mathrm{E3}}$ yields $\nu_{\mathrm{E2}} = 688\,358\,979\,309\,308.22(9)$~Hz.

The measurements presented above provide a stringent test of LPI for the specific pairs of compared atomic transition frequencies. Based on a general parametrization \cite{fla09}, a possible variation $dF$ of any atomic transition frequency can be related to variations of three constants: the fine structure constant $\alpha$, the proton-to-electron mass ratio $\mu=m_p/m_e$ and the ratio between the average quark mass and the quantum chromodynamic scale $X_q=m_q/\Lambda_\mathrm{QCD}$, according to
\begin{equation}
    \frac{1}{F} dF = \kappa_\alpha \frac{1}{\alpha} d\alpha + \kappa_\mu \frac{1}{\mu} d\mu + \kappa_q \frac{1}{X_q} dX_q,
\end{equation}
where $\kappa_\alpha$, $\kappa_\mu$ and $\kappa_q$ describe the sensitivity of the transition frequency $F$ to the constants.
A sensitivity of $\kappa_\mu=-1$ results from the nuclear magneton for all magnetic hyperfine splittings, while $\kappa_q$ varies for hyperfine transitions of different atomic species. 
Both, hyperfine and optical electronic transitions, obtain a sensitivity for variations of $\alpha$ from relativistic contributions to the transition energies. 
The factors $\kappa_\alpha$ and $\kappa_q$ can be determined with numerical many-body calculations \cite{fla09,din09}.

The ratio $R_{\mathrm{E3,E2}}$ of the two $^{171}$Yb$^+$ transition frequencies is sensitive to variations of $\alpha$ only and an explicit sensitivity $\kappa_\alpha(R_{\mathrm{E3,E2}}) = \kappa_\alpha(\mathrm{E3})-\kappa_\alpha(\mathrm{E2})=-6.95 $ has been calculated \cite{fla09}. 
The data presented in Fig.~\ref{graph:RatiovsTime} show a fractional temporal drift $(1/R_{\mathrm{E3,E2}})(dR_{\mathrm{E3,E2}}/dt)=-6.8(7.5)\times10^{-18}/\mathrm{yr}$ of the frequency ratio and allows us to infer a potential temporal variation $(1/\alpha) (d\alpha/dt)=1.0(1.1)\times10^{-18}/\mathrm{yr}$, compatible with zero. 
This constitutes an improvement over the previous limits by more than one order of magnitude \cite{ros08,god14,hun14}. This finding is summarized in Table~\ref{tab:limits} together with limits obtained below.
\begin{table}
    \centering
    \begin{ruledtabular}
    \begin{tabular}{c|c|c|c|c}
    \multirow{2}{*}{$x$}    & \multicolumn{2}{c|}{$(1/x)(dx/dt) \left(10^{-18} / \mathrm{yr} \right) $}    
         & \multicolumn{2}{c}{$(c^2/x) (dx/d\Phi)$ $\left(10^{-8}\right) $}\\ 
                                &this work & previous &this work & previous \\\hline
    $R_{\mathrm{E3,E2}}$        & -6.8(7.5)&                            & -9.7(7.7) &                       \\
    $\alpha$                    & 1.0(1.1) & -16(23) \cite{ros08}       & 1.4(1.1)  & -5.3(10) \cite{dzu17} \\
    $R_{\mathrm{E3,Cs}}$        & -31(34)  & 20(410) \cite{hun14}       & -4(44)    &                       \\
    $\mu$                       & -8(36)   & $53(65)$ \cite{mcg19}      & 7(45)     & 35(59) \cite{sch20d}  \\
    \end{tabular}
    \end{ruledtabular}
    \caption{Limits on a violation of local position invariance of atomic frequency ratios and fundamental constants for a linear temporal variation and a coupling to changes of the gravitational potential $\Phi$. Listed quantities $x$ are the ratio $R_{\mathrm{E3,E2}}$ of the E3 and E2 transitions frequencies of $^{171}$Yb$^+$, the ratio $R_{\mathrm{E3,Cs}}$ of the E3 transition frequency and the Cs hyperfine splitting frequency, the fine structure constant $\alpha$ and the proton-to-electron mass ratio $\mu=m_p/m_e$. The speed of light is denoted with $c$. }
    \label{tab:limits}
\end{table}

In comparisons of two hyperfine transition frequencies of different atoms the sensitivity to $\mu$ is common-mode and a dependence only on $\alpha$ and $X_q$ remains. 
From measurements over 14 years with microwave fountain clocks with $^{133}$Cs and $^{87}$Rb atoms, and the previous stringent constraint on $d\alpha/dt$ \cite{ros08}, a potential temporal variation in $X_q$ has been restricted to $\kappa_q(\mathrm{Cs}) (1/X_q) (dX_q/dt)= 0.14(9)\times10^{-16}/\mathrm{yr}$ in Ref.~\cite{gue12}. 
Comparisons between frequencies of electronic transitions and hyperfine transitions, such as absolute frequency measurements of optical transitions using caesium clocks, provide a non-zero sensitivity on $d\alpha/dt$, $d\mu/dt$ and $dX_q/dt$. 
For the measurements of $\nu_{\mathrm{E3}}$ with caesium fountain clocks shown in Fig.~\ref{graph:FreqvsTime}, we find a relative temporal drift $(1/R_{\mathrm{E3,Cs}}) (dR_{\mathrm{E3,Cs}}/dt) = -3.1(3.4)\times 10^{-17}/\mathrm{yr}$. 
Using the restrictions on $(1/\alpha) (d\alpha/dt)$ from the $\nu_{\mathrm{E3}}/\nu_{\mathrm{E2}}$ measurements and on $(1/X_q) (dX_q/dt)$ from Ref~\cite{gue12}, we find $(1/\mu) (d\mu/dt) = -8(36)\times10^{-18}/\mathrm{yr}$. 
This improves the uncertainty in the limits obtained in Refs~\cite{mcg19,sch20d} by about a factor 2. 

In addition to the search for a steady temporal drift, we can use the data presented above to search for a coupling of the fundamental  constants $\alpha$ and $\mu$ to gravity in an analysis of gravitational red-shift \cite{wil14}. 
The variation of the Sun's gravitational potential on Earth $\Phi (t) $ due to the ellipticity of the Earth orbit can be approximated by $\Delta \Phi \cos(2\pi (t-t_p)/T_a)$ with $\Delta \Phi/c^2 \approx 1.65\times 10^{-10}$, $c$ the speed of light, $t_p$ the time of the perihelion 2018 and $T_a$ the anomalistic year. 
Similarly to the search for a temporal variation, we start with data of the optical frequency ratio and investigate for a potential coupling to gravity by non-linear least square fitting of $A\cos(2\pi (t-t_p)/T_a) + B$ to the data shown in Fig.~\ref{graph:RatiovsTime} with $A$ and $B$ as free parameters. We find a relative amplitude $A/R_{\mathrm{E3,E2}}=-16(13)\times 10^{-18}$.  
Because of the large sensitivity $\kappa_\alpha$, the result limits a potential coupling of $\alpha$ to gravity to $(c^2/\alpha) (d\alpha/d\Phi) = 14(11)\times 10^{-9}$ and improves previous limits from Ref.~\cite{dzu17} by one order of magnitude. 

We combine this result with the obtained oscillation amplitude $c^2A/\Delta \Phi = 22(25)\times 10^{-8}$ found in a comparison between caesium clocks and H-masers \cite{ash18} which yields $(c^2/X_q) (dX_q/d\Phi) = -21(23)\times 10^{-7}$.
For the data shown in Fig.~\ref{graph:FreqvsTime}, we find $A/\nu_{\mathrm{E3}}=-7(72)\times 10^{-18}$ when searching for an oscillation induced by the annual variation of the gravitational potential. 
In combination with the dependence on $\alpha$ and $X_q$ given above, we find $(c^2/\mu) (d\mu/d\Phi) = 7(45)\times 10^{-8}$ slightly improving the limit obtained in \cite{sch20d}.

The presented results on LPI violating parameters, summarized in Table~\ref{tab:limits}, are all consistent with zero to within less than 1.3 standard deviations and support the validity of this fundamental assumption of general relativity in the solar system in the present epoch. 
The large potential for improved searches promised by the small fractional uncertainty of the E3 clock and its high sensitivity to changes of $\alpha$ can be readily explored in comparisons with other high performance clocks.
This also opens up new possibilities in searches for ultralight scalar dark matter \cite{arv15,rob20}. 

We thank Thomas Legero, Erik Benkler and Uwe Sterr for providing the ultrastable laser reference Si-2 and Tara Cubel Liebisch for a critical reading of the manuscript.
This project has received funding from the EMPIR programme co-financed by the Participating States and from the European Union’s Horizon 2020 research and innovation programme.
It has been supported by the EMPIR projects 17FUN07 "Coulomb Crystals for Clocks" and 18SIB05 "Robust Optical Clocks for International Timescales". 
Furthermore this work has been supported by the Max-Planck-RIKEN-PTB-Center for Time, Constants and Fundamental Symmetries, and by the Deutsche Forschungsgemeinschaft (DFG, German Research Foundation) under CRC~1227 DQ-\textit{mat} within project B02. C.~Sanner thanks the Alexander von Humboldt Foundation for support.

\bibliographystyle{apsrev4-2}

\begin{thebibliography}{41}%
\makeatletter
\providecommand \@ifxundefined [1]{%
 \@ifx{#1\undefined}
}%
\providecommand \@ifnum [1]{%
 \ifnum #1\expandafter \@firstoftwo
 \else \expandafter \@secondoftwo
 \fi
}%
\providecommand \@ifx [1]{%
 \ifx #1\expandafter \@firstoftwo
 \else \expandafter \@secondoftwo
 \fi
}%
\providecommand \natexlab [1]{#1}%
\providecommand \enquote  [1]{``#1''}%
\providecommand \bibnamefont  [1]{#1}%
\providecommand \bibfnamefont [1]{#1}%
\providecommand \citenamefont [1]{#1}%
\providecommand \href@noop [0]{\@secondoftwo}%
\providecommand \href [0]{\begingroup \@sanitize@url \@href}%
\providecommand \@href[1]{\@@startlink{#1}\@@href}%
\providecommand \@@href[1]{\endgroup#1\@@endlink}%
\providecommand \@sanitize@url [0]{\catcode `\\12\catcode `\$12\catcode
  `\&12\catcode `\#12\catcode `\^12\catcode `\_12\catcode `\%12\relax}%
\providecommand \@@startlink[1]{}%
\providecommand \@@endlink[0]{}%
\providecommand \url  [0]{\begingroup\@sanitize@url \@url }%
\providecommand \@url [1]{\endgroup\@href {#1}{\urlprefix }}%
\providecommand \urlprefix  [0]{URL }%
\providecommand \Eprint [0]{\href }%
\providecommand \doibase [0]{https://doi.org/}%
\providecommand \selectlanguage [0]{\@gobble}%
\providecommand \bibinfo  [0]{\@secondoftwo}%
\providecommand \bibfield  [0]{\@secondoftwo}%
\providecommand \translation [1]{[#1]}%
\providecommand \BibitemOpen [0]{}%
\providecommand \bibitemStop [0]{}%
\providecommand \bibitemNoStop [0]{.\EOS\space}%
\providecommand \EOS [0]{\spacefactor3000\relax}%
\providecommand \BibitemShut  [1]{\csname bibitem#1\endcsname}%
\let\auto@bib@innerbib\@empty
\bibitem [{\citenamefont {Safronova}\ \emph {et~al.}(2018)\citenamefont
  {Safronova}, \citenamefont {Budker}, \citenamefont {DeMille}, \citenamefont
  {Jackson~Kimball}, \citenamefont {Derevianko},\ and\ \citenamefont
  {W.~Clark}}]{saf18a}%
  \BibitemOpen
  \bibfield  {author} {\bibinfo {author} {\bibfnamefont {M.~S.}~\bibnamefont
  {Safronova}}, \bibinfo {author} {\bibfnamefont {D.}~\bibnamefont {Budker}},
  \bibinfo {author} {\bibfnamefont {D.}~\bibnamefont {DeMille}}, \bibinfo
  {author} {\bibfnamefont {D.~F.}\ \bibnamefont {Jackson~Kimball}}, \bibinfo
  {author} {\bibfnamefont {A.}~\bibnamefont {Derevianko}},\ and\ \bibinfo
  {author} {\bibfnamefont {C.}~\bibnamefont {W.~Clark}},\ }\href
  {https://doi.org/10.1103/RevModPhys.90.025008} {\bibfield  {journal}
  {\bibinfo  {journal} {Rev. Mod. Phys.}\ }\textbf {\bibinfo {volume} {90}},\
  \bibinfo {pages} {025008} (\bibinfo {year} {2018})}\BibitemShut {NoStop}%
\bibitem [{\citenamefont {Will}(2014)}]{wil14}%
  \BibitemOpen
  \bibfield  {author} {\bibinfo {author} {\bibfnamefont {C.~M.}\ \bibnamefont
  {Will}},\ }\href {https://doi.org/10.12942/lrr-2014-4} {\bibfield  {journal}
  {\bibinfo  {journal} {Living Rev. Relativity}\ }\textbf {\bibinfo {volume}
  {9}},\ \bibinfo {pages} {3} (\bibinfo {year} {2014})},\ \bibinfo {note}
  {[Online Article]: cited 2014-07-01, updated in 2014}\BibitemShut {NoStop}%
\bibitem [{\citenamefont {Martins}(2017)}]{mar17}%
  \BibitemOpen
  \bibfield  {author} {\bibinfo {author} {\bibfnamefont {C.~J. A.~P.}\
  \bibnamefont {Martins}},\ }\href {https://doi.org/10.1088/1361-6633/aa860e}
  {\bibfield  {journal} {\bibinfo  {journal} {Reports on Progress in Physics}\
  }\textbf {\bibinfo {volume} {80}},\ \bibinfo {pages} {126902} (\bibinfo
  {year} {2017})}\BibitemShut {NoStop}%
\bibitem [{\citenamefont {Wilczynska}\ \emph {et~al.}(2020)\citenamefont
  {Wilczynska}, \citenamefont {Webb}, \citenamefont {Bainbridge}, \citenamefont
  {Barrow}, \citenamefont {Bosman}, \citenamefont {Carswell}, \citenamefont
  {D{\k a}browski}, \citenamefont {Dumont}, \citenamefont {Lee}, \citenamefont
  {Leite}, \citenamefont {Leszczy{\'n}ska}, \citenamefont {Liske},
  \citenamefont {Marosek}, \citenamefont {Martins}, \citenamefont
  {Milakovi{\'c}}, \citenamefont {Molaro},\ and\ \citenamefont
  {Pasquini}}]{wil20}%
  \BibitemOpen
  \bibfield  {author} {\bibinfo {author} {\bibfnamefont {M.~R.}\ \bibnamefont
  {Wilczynska}}, \bibinfo {author} {\bibfnamefont {J.~K.}\ \bibnamefont
  {Webb}}, \bibinfo {author} {\bibfnamefont {M.}~\bibnamefont {Bainbridge}},
  \bibinfo {author} {\bibfnamefont {J.~D.}\ \bibnamefont {Barrow}}, \bibinfo
  {author} {\bibfnamefont {S.~E.~I.}\ \bibnamefont {Bosman}}, \bibinfo {author}
  {\bibfnamefont {R.~F.}\ \bibnamefont {Carswell}}, \bibinfo {author}
  {\bibfnamefont {M.~P.}\ \bibnamefont {D{a}browski}}, \bibinfo {author}
  {\bibfnamefont {V.}~\bibnamefont {Dumont}}, \bibinfo {author} {\bibfnamefont
  {C.-C.}\ \bibnamefont {Lee}}, \bibinfo {author} {\bibfnamefont {A.~C.}\
  \bibnamefont {Leite}}, \bibinfo {author} {\bibfnamefont {K.}~\bibnamefont
  {Leszczy{\'n}ska}}, \bibinfo {author} {\bibfnamefont {J.}~\bibnamefont
  {Liske}}, \bibinfo {author} {\bibfnamefont {K.}~\bibnamefont {Marosek}},
  \bibinfo {author} {\bibfnamefont {C.~J. A.~P.}\ \bibnamefont {Martins}},
  \bibinfo {author} {\bibfnamefont {D.}~\bibnamefont {Milakovi{\'c}}}, \bibinfo
  {author} {\bibfnamefont {P.}~\bibnamefont {Molaro}},\ and\ \bibinfo {author}
  {\bibfnamefont {L.}~\bibnamefont {Pasquini}},\ }\bibfield  {journal}
  {\bibinfo  {journal} {Science Advances}\ }\textbf {\bibinfo {volume} {6}},\
  \href {https://doi.org/10.1126/sciadv.aay9672} {10.1126/sciadv.aay9672}
  (\bibinfo {year} {2020})\BibitemShut {NoStop}%
\bibitem [{\citenamefont {Rosenband}\ \emph {et~al.}(2008)\citenamefont
  {Rosenband}, \citenamefont {Hume}, \citenamefont {Schmidt}, \citenamefont
  {Chou}, \citenamefont {Brusch}, \citenamefont {Lorini}, \citenamefont
  {Oskay}, \citenamefont {Drullinger}, \citenamefont {Fortier}, \citenamefont
  {Stalnaker}, \citenamefont {Diddams}, \citenamefont {Swann}, \citenamefont
  {Newbury}, \citenamefont {Itano}, \citenamefont {Wineland},\ and\
  \citenamefont {Bergquist}}]{ros08}%
  \BibitemOpen
  \bibfield  {author} {\bibinfo {author} {\bibfnamefont {T.}~\bibnamefont
  {Rosenband}}, \bibinfo {author} {\bibfnamefont {D.~B.}\ \bibnamefont {Hume}},
  \bibinfo {author} {\bibfnamefont {P.~O.}\ \bibnamefont {Schmidt}}, \bibinfo
  {author} {\bibfnamefont {C.~W.}\ \bibnamefont {Chou}}, \bibinfo {author}
  {\bibfnamefont {A.}~\bibnamefont {Brusch}}, \bibinfo {author} {\bibfnamefont
  {L.}~\bibnamefont {Lorini}}, \bibinfo {author} {\bibfnamefont {W.~H.}\
  \bibnamefont {Oskay}}, \bibinfo {author} {\bibfnamefont {R.~E.}\ \bibnamefont
  {Drullinger}}, \bibinfo {author} {\bibfnamefont {T.~M.}\ \bibnamefont
  {Fortier}}, \bibinfo {author} {\bibfnamefont {J.~E.}\ \bibnamefont
  {Stalnaker}}, \bibinfo {author} {\bibfnamefont {S.~A.}\ \bibnamefont
  {Diddams}}, \bibinfo {author} {\bibfnamefont {W.~C.}\ \bibnamefont {Swann}},
  \bibinfo {author} {\bibfnamefont {N.~R.}\ \bibnamefont {Newbury}}, \bibinfo
  {author} {\bibfnamefont {W.~M.}\ \bibnamefont {Itano}}, \bibinfo {author}
  {\bibfnamefont {D.~J.}\ \bibnamefont {Wineland}},\ and\ \bibinfo {author}
  {\bibfnamefont {J.~C.}\ \bibnamefont {Bergquist}},\ }\href
  {https://doi.org/10.1126/science.1154622} {\bibfield  {journal} {\bibinfo
  {journal} {Science}\ }\textbf {\bibinfo {volume} {319}},\ \bibinfo {pages}
  {1808} (\bibinfo {year} {2008})}\BibitemShut {NoStop}%
\bibitem [{\citenamefont {Godun}\ \emph {et~al.}(2014)\citenamefont {Godun},
  \citenamefont {Nisbet-Jones}, \citenamefont {Jones}, \citenamefont {King},
  \citenamefont {Johnson}, \citenamefont {Margolis}, \citenamefont {Szymaniec},
  \citenamefont {Lea}, \citenamefont {Bongs},\ and\ \citenamefont
  {Gill}}]{god14}%
  \BibitemOpen
  \bibfield  {author} {\bibinfo {author} {\bibfnamefont {R.~M.}\ \bibnamefont
  {Godun}}, \bibinfo {author} {\bibfnamefont {P.~B.~R.}\ \bibnamefont
  {Nisbet-Jones}}, \bibinfo {author} {\bibfnamefont {J.~M.}\ \bibnamefont
  {Jones}}, \bibinfo {author} {\bibfnamefont {S.~A.}\ \bibnamefont {King}},
  \bibinfo {author} {\bibfnamefont {L.~A.~M.}\ \bibnamefont {Johnson}},
  \bibinfo {author} {\bibfnamefont {H.~S.}\ \bibnamefont {Margolis}}, \bibinfo
  {author} {\bibfnamefont {K.}~\bibnamefont {Szymaniec}}, \bibinfo {author}
  {\bibfnamefont {S.~N.}\ \bibnamefont {Lea}}, \bibinfo {author} {\bibfnamefont
  {K.}~\bibnamefont {Bongs}},\ and\ \bibinfo {author} {\bibfnamefont
  {P.}~\bibnamefont {Gill}},\ }\href
  {https://doi.org/10.1103/PhysRevLett.113.210801} {\bibfield  {journal}
  {\bibinfo  {journal} {Phys. Rev. Lett.}\ }\textbf {\bibinfo {volume} {113}},\
  \bibinfo {pages} {210801} (\bibinfo {year} {2014})}\BibitemShut {NoStop}%
\bibitem [{\citenamefont {Huntemann}\ \emph {et~al.}(2014)\citenamefont
  {Huntemann}, \citenamefont {Lipphardt}, \citenamefont {Tamm}, \citenamefont
  {Gerginov}, \citenamefont {Weyers},\ and\ \citenamefont {Peik}}]{hun14}%
  \BibitemOpen
  \bibfield  {author} {\bibinfo {author} {\bibfnamefont {N.}~\bibnamefont
  {Huntemann}}, \bibinfo {author} {\bibfnamefont {B.}~\bibnamefont
  {Lipphardt}}, \bibinfo {author} {\bibfnamefont {C.}~\bibnamefont {Tamm}},
  \bibinfo {author} {\bibfnamefont {V.}~\bibnamefont {Gerginov}}, \bibinfo
  {author} {\bibfnamefont {S.}~\bibnamefont {Weyers}},\ and\ \bibinfo {author}
  {\bibfnamefont {E.}~\bibnamefont {Peik}},\ }\href
  {https://doi.org/10.1103/PhysRevLett.113.210802} {\bibfield  {journal}
  {\bibinfo  {journal} {Phys. Rev. Lett.}\ }\textbf {\bibinfo {volume} {113}},\
  \bibinfo {pages} {210802} (\bibinfo {year} {2014})}\BibitemShut {NoStop}%
\bibitem [{\citenamefont {Gu{\'e}na}\ \emph {et~al.}(2012)\citenamefont
  {Gu{\'e}na}, \citenamefont {Abgrall}, \citenamefont {Rovera}, \citenamefont
  {Rosenbusch}, \citenamefont {Tobar}, \citenamefont {Laurent}, \citenamefont
  {Clairon},\ and\ \citenamefont {Bize}}]{gue12}%
  \BibitemOpen
  \bibfield  {author} {\bibinfo {author} {\bibfnamefont {J.}~\bibnamefont
  {Gu{\'e}na}}, \bibinfo {author} {\bibfnamefont {M.}~\bibnamefont {Abgrall}},
  \bibinfo {author} {\bibfnamefont {D.}~\bibnamefont {Rovera}}, \bibinfo
  {author} {\bibfnamefont {P.}~\bibnamefont {Rosenbusch}}, \bibinfo {author}
  {\bibfnamefont {M.~E.}\ \bibnamefont {Tobar}}, \bibinfo {author}
  {\bibfnamefont {P.}~\bibnamefont {Laurent}}, \bibinfo {author} {\bibfnamefont
  {A.}~\bibnamefont {Clairon}},\ and\ \bibinfo {author} {\bibfnamefont
  {S.}~\bibnamefont {Bize}},\ }\href
  {https://doi.org/10.1103/PhysRevLett.109.080801} {\bibfield  {journal}
  {\bibinfo  {journal} {Phys. Rev. Lett.}\ }\textbf {\bibinfo {volume} {109}},\
  \bibinfo {pages} {080801} (\bibinfo {year} {2012})}\BibitemShut {NoStop}%
\bibitem [{\citenamefont {Ashby}\ \emph {et~al.}(2018)\citenamefont {Ashby},
  \citenamefont {Parker},\ and\ \citenamefont {Patla}}]{ash18}%
  \BibitemOpen
  \bibfield  {author} {\bibinfo {author} {\bibfnamefont {N.}~\bibnamefont
  {Ashby}}, \bibinfo {author} {\bibfnamefont {T.~E.}\ \bibnamefont {Parker}},\
  and\ \bibinfo {author} {\bibfnamefont {B.~R.}\ \bibnamefont {Patla}},\ }\href
  {https://doi.org/10.1038/s41567-018-0156-2} {\bibfield  {journal} {\bibinfo
  {journal} {Nature Physics}\ }\textbf {\bibinfo {volume} {14}},\ \bibinfo
  {pages} {822 } (\bibinfo {year} {2018})}\BibitemShut {NoStop}%
\bibitem [{\citenamefont {McGrew}\ \emph {et~al.}(2018)\citenamefont {McGrew},
  \citenamefont {Zhang}, \citenamefont {Fasano}, \citenamefont {Sch\"affer},
  \citenamefont {Beloy}, \citenamefont {Nicolodi}, \citenamefont {Brown},
  \citenamefont {Hinkley}, \citenamefont {Milan}, \citenamefont {Schioppo},
  \citenamefont {Yoon},\ and\ \citenamefont {Ludlow}}]{mcg18}%
  \BibitemOpen
  \bibfield  {author} {\bibinfo {author} {\bibfnamefont {W.~F.}\ \bibnamefont
  {McGrew}}, \bibinfo {author} {\bibfnamefont {X.}~\bibnamefont {Zhang}},
  \bibinfo {author} {\bibfnamefont {R.~J.}\ \bibnamefont {Fasano}}, \bibinfo
  {author} {\bibfnamefont {S.~A.}\ \bibnamefont {Sch\"affer}}, \bibinfo
  {author} {\bibfnamefont {K.}~\bibnamefont {Beloy}}, \bibinfo {author}
  {\bibfnamefont {D.}~\bibnamefont {Nicolodi}}, \bibinfo {author}
  {\bibfnamefont {R.~C.}\ \bibnamefont {Brown}}, \bibinfo {author}
  {\bibfnamefont {N.}~\bibnamefont {Hinkley}}, \bibinfo {author} {\bibfnamefont
  {G.}~\bibnamefont {Milan}}, \bibinfo {author} {\bibfnamefont
  {M.}~\bibnamefont {Schioppo}}, \bibinfo {author} {\bibfnamefont {T.~H.}\
  \bibnamefont {Yoon}},\ and\ \bibinfo {author} {\bibfnamefont {A.~D.}\
  \bibnamefont {Ludlow}},\ }\href {https://doi.org/10.1038/s41586-018-0738-2}
  {\bibfield  {journal} {\bibinfo  {journal} {Nature}\ }\textbf {\bibinfo
  {volume} {564}},\ \bibinfo {pages} {87} (\bibinfo {year} {2018})}\BibitemShut
  {NoStop}%
\bibitem [{\citenamefont {Bothwell}\ \emph {et~al.}(2019)\citenamefont
  {Bothwell}, \citenamefont {Kedar}, \citenamefont {Oelker}, \citenamefont
  {Robinson}, \citenamefont {Bromley}, \citenamefont {Tew}, \citenamefont
  {Ye},\ and\ \citenamefont {Kennedy}}]{bot19}%
  \BibitemOpen
  \bibfield  {author} {\bibinfo {author} {\bibfnamefont {T.}~\bibnamefont
  {Bothwell}}, \bibinfo {author} {\bibfnamefont {D.}~\bibnamefont {Kedar}},
  \bibinfo {author} {\bibfnamefont {E.}~\bibnamefont {Oelker}}, \bibinfo
  {author} {\bibfnamefont {J.~M.}\ \bibnamefont {Robinson}}, \bibinfo {author}
  {\bibfnamefont {S.~L.}\ \bibnamefont {Bromley}}, \bibinfo {author}
  {\bibfnamefont {W.~L.}\ \bibnamefont {Tew}}, \bibinfo {author} {\bibfnamefont
  {J.}~\bibnamefont {Ye}},\ and\ \bibinfo {author} {\bibfnamefont {C.~J.}\
  \bibnamefont {Kennedy}},\ }\href {https://doi.org/10.1088/1681-7575/ab4089}
  {\bibfield  {journal} {\bibinfo  {journal} {Metrologia}\ }\textbf {\bibinfo
  {volume} {56}},\ \bibinfo {pages} {065004} (\bibinfo {year}
  {2019})}\BibitemShut {NoStop}%
\bibitem [{\citenamefont {Sanner}\ \emph {et~al.}(2019)\citenamefont {Sanner},
  \citenamefont {Huntemann}, \citenamefont {Lange}, \citenamefont {Tamm},
  \citenamefont {Peik}, \citenamefont {Safronova},\ and\ \citenamefont
  {Porsev}}]{san19}%
  \BibitemOpen
  \bibfield  {author} {\bibinfo {author} {\bibfnamefont {C.}~\bibnamefont
  {Sanner}}, \bibinfo {author} {\bibfnamefont {N.}~\bibnamefont {Huntemann}},
  \bibinfo {author} {\bibfnamefont {R.}~\bibnamefont {Lange}}, \bibinfo
  {author} {\bibfnamefont {C.}~\bibnamefont {Tamm}}, \bibinfo {author}
  {\bibfnamefont {E.}~\bibnamefont {Peik}}, \bibinfo {author} {\bibfnamefont
  {M.~S.}\ \bibnamefont {Safronova}},\ and\ \bibinfo {author} {\bibfnamefont
  {S.~G.}\ \bibnamefont {Porsev}},\ }\href
  {https://doi.org/10.1038/s41586-019-0972-2} {\bibfield  {journal} {\bibinfo
  {journal} {Nature}\ }\textbf {\bibinfo {volume} {567}},\ \bibinfo {pages}
  {204} (\bibinfo {year} {2019})}\BibitemShut {NoStop}%
\bibitem [{\citenamefont {Brewer}\ \emph {et~al.}(2019)\citenamefont {Brewer},
  \citenamefont {Chen}, \citenamefont {Hankin}, \citenamefont {Clements},
  \citenamefont {Chou}, \citenamefont {Wineland}, \citenamefont {Hume},\ and\
  \citenamefont {Leibrandt}}]{bre19}%
  \BibitemOpen
  \bibfield  {author} {\bibinfo {author} {\bibfnamefont {S.~M.}\ \bibnamefont
  {Brewer}}, \bibinfo {author} {\bibfnamefont {J.-S.}\ \bibnamefont {Chen}},
  \bibinfo {author} {\bibfnamefont {A.~M.}\ \bibnamefont {Hankin}}, \bibinfo
  {author} {\bibfnamefont {E.~R.}\ \bibnamefont {Clements}}, \bibinfo {author}
  {\bibfnamefont {C.~W.}\ \bibnamefont {Chou}}, \bibinfo {author}
  {\bibfnamefont {D.~J.}\ \bibnamefont {Wineland}}, \bibinfo {author}
  {\bibfnamefont {D.~B.}\ \bibnamefont {Hume}},\ and\ \bibinfo {author}
  {\bibfnamefont {D.~R.}\ \bibnamefont {Leibrandt}},\ }\href
  {https://doi.org/10.1103/PhysRevLett.123.033201} {\bibfield  {journal}
  {\bibinfo  {journal} {Phys. Rev. Lett.}\ }\textbf {\bibinfo {volume} {123}},\
  \bibinfo {pages} {033201} (\bibinfo {year} {2019})}\BibitemShut {NoStop}%
\bibitem [{\citenamefont {Kozlov}\ \emph {et~al.}(2018)\citenamefont {Kozlov},
  \citenamefont {Safronova}, \citenamefont {Crespo L\'opez-Urrutia},\ and\
  \citenamefont {Schmidt}}]{koz18}%
  \BibitemOpen
  \bibfield  {author} {\bibinfo {author} {\bibfnamefont {M.~G.}\ \bibnamefont
  {Kozlov}}, \bibinfo {author} {\bibfnamefont {M.~S.}\ \bibnamefont
  {Safronova}}, \bibinfo {author} {\bibfnamefont {J.~R.}\ \bibnamefont {Crespo
  L\'opez-Urrutia}},\ and\ \bibinfo {author} {\bibfnamefont {P.~O.}\
  \bibnamefont {Schmidt}},\ }\href
  {https://doi.org/10.1103/RevModPhys.90.045005} {\bibfield  {journal}
  {\bibinfo  {journal} {Rev. Mod. Phys.}\ }\textbf {\bibinfo {volume} {90}},\
  \bibinfo {pages} {045005} (\bibinfo {year} {2018})}\BibitemShut {NoStop}%
\bibitem [{\citenamefont {Flambaum}(2006)}]{fla06}%
  \BibitemOpen
  \bibfield  {author} {\bibinfo {author} {\bibfnamefont {V.~V.}\ \bibnamefont
  {Flambaum}},\ }\href {http://link.aps.org/abstract/PRL/v97/e092502}
  {\bibfield  {journal} {\bibinfo  {journal} {Phys. Rev. Lett.}\ }\textbf
  {\bibinfo {volume} {97}},\ \bibinfo {pages} {092502} (\bibinfo {year}
  {2006})}\BibitemShut {NoStop}%
\bibitem [{\citenamefont {Thielking}\ \emph {et~al.}(2018)\citenamefont
  {Thielking}, \citenamefont {Okhapkin}, \citenamefont {G{\l}owacki},
  \citenamefont {Meier}, \citenamefont {von~der Wense}, \citenamefont
  {Seiferle}, \citenamefont {D{\"u}llmann}, \citenamefont {Thirolf},\ and\
  \citenamefont {Peik}}]{thi18}%
  \BibitemOpen
  \bibfield  {author} {\bibinfo {author} {\bibfnamefont {J.}~\bibnamefont
  {Thielking}}, \bibinfo {author} {\bibfnamefont {M.~V.}\ \bibnamefont
  {Okhapkin}}, \bibinfo {author} {\bibfnamefont {P.}~\bibnamefont
  {G{\l}owacki}}, \bibinfo {author} {\bibfnamefont {D.~M.}\ \bibnamefont
  {Meier}}, \bibinfo {author} {\bibfnamefont {L.}~\bibnamefont {von~der
  Wense}}, \bibinfo {author} {\bibfnamefont {B.}~\bibnamefont {Seiferle}},
  \bibinfo {author} {\bibfnamefont {C.~E.}\ \bibnamefont {D{\"u}llmann}},
  \bibinfo {author} {\bibfnamefont {P.~G.}\ \bibnamefont {Thirolf}},\ and\
  \bibinfo {author} {\bibfnamefont {E.}~\bibnamefont {Peik}},\ }\href
  {https://doi.org/10.1038/s41586-018-0011-8} {\bibfield  {journal} {\bibinfo
  {journal} {Nature}\ }\textbf {\bibinfo {volume} {556}},\ \bibinfo {pages}
  {321} (\bibinfo {year} {2018})}\BibitemShut {NoStop}%
\bibitem [{\citenamefont {Roberts}\ \emph {et~al.}(2000)\citenamefont
  {Roberts}, \citenamefont {Taylor}, \citenamefont {Barwood}, \citenamefont
  {Rowley},\ and\ \citenamefont {Gill}}]{rob00}%
  \BibitemOpen
  \bibfield  {author} {\bibinfo {author} {\bibfnamefont {M.}~\bibnamefont
  {Roberts}}, \bibinfo {author} {\bibfnamefont {P.}~\bibnamefont {Taylor}},
  \bibinfo {author} {\bibfnamefont {G.~P.}\ \bibnamefont {Barwood}}, \bibinfo
  {author} {\bibfnamefont {W.~R.~C.}\ \bibnamefont {Rowley}},\ and\ \bibinfo
  {author} {\bibfnamefont {P.}~\bibnamefont {Gill}},\ }\href
  {https://doi.org/10.1103/PhysRevA.62.020501} {\bibfield  {journal} {\bibinfo
  {journal} {Phys. Rev. A}\ }\textbf {\bibinfo {volume} {62}},\ \bibinfo
  {pages} {020501(R)} (\bibinfo {year} {2000})}\BibitemShut {NoStop}%
\bibitem [{\citenamefont {Matei}\ \emph {et~al.}(2017)\citenamefont {Matei},
  \citenamefont {Legero}, \citenamefont {H\"afner}, \citenamefont {Grebing},
  \citenamefont {Weyrich}, \citenamefont {Zhang}, \citenamefont {Sonderhouse},
  \citenamefont {Robinson}, \citenamefont {Ye}, \citenamefont {Riehle},\ and\
  \citenamefont {Sterr}}]{mat17a}%
  \BibitemOpen
  \bibfield  {author} {\bibinfo {author} {\bibfnamefont {D.~G.}\ \bibnamefont
  {Matei}}, \bibinfo {author} {\bibfnamefont {T.}~\bibnamefont {Legero}},
  \bibinfo {author} {\bibfnamefont {S.}~\bibnamefont {H\"afner}}, \bibinfo
  {author} {\bibfnamefont {C.}~\bibnamefont {Grebing}}, \bibinfo {author}
  {\bibfnamefont {R.}~\bibnamefont {Weyrich}}, \bibinfo {author} {\bibfnamefont
  {W.}~\bibnamefont {Zhang}}, \bibinfo {author} {\bibfnamefont
  {L.}~\bibnamefont {Sonderhouse}}, \bibinfo {author} {\bibfnamefont {J.~M.}\
  \bibnamefont {Robinson}}, \bibinfo {author} {\bibfnamefont {J.}~\bibnamefont
  {Ye}}, \bibinfo {author} {\bibfnamefont {F.}~\bibnamefont {Riehle}},\ and\
  \bibinfo {author} {\bibfnamefont {U.}~\bibnamefont {Sterr}},\ }\href
  {https://doi.org/10.1103/PhysRevLett.118.263202} {\bibfield  {journal}
  {\bibinfo  {journal} {Phys. Rev. Lett.}\ }\textbf {\bibinfo {volume} {118}},\
  \bibinfo {pages} {263202} (\bibinfo {year} {2017})}\BibitemShut {NoStop}%
\bibitem [{\citenamefont {Flambaum}\ and\ \citenamefont {Dzuba}(2009)}]{fla09}%
  \BibitemOpen
  \bibfield  {author} {\bibinfo {author} {\bibfnamefont {V.~V.}\ \bibnamefont
  {Flambaum}}\ and\ \bibinfo {author} {\bibfnamefont {V.~A.}\ \bibnamefont
  {Dzuba}},\ }\href {https://doi.org/10.1139/P08-072} {\bibfield  {journal}
  {\bibinfo  {journal} {Can. J. Phys.}\ }\textbf {\bibinfo {volume} {87}},\
  \bibinfo {pages} {25} (\bibinfo {year} {2009})}\BibitemShut {NoStop}%
\bibitem [{\citenamefont {Ejtemaee}\ \emph {et~al.}(2010)\citenamefont
  {Ejtemaee}, \citenamefont {Thomas},\ and\ \citenamefont {Haljan}}]{ejt10}%
  \BibitemOpen
  \bibfield  {author} {\bibinfo {author} {\bibfnamefont {S.}~\bibnamefont
  {Ejtemaee}}, \bibinfo {author} {\bibfnamefont {R.}~\bibnamefont {Thomas}},\
  and\ \bibinfo {author} {\bibfnamefont {P.~C.}\ \bibnamefont {Haljan}},\
  }\href {https://doi.org/10.1103/PhysRevA.82.063419} {\bibfield  {journal}
  {\bibinfo  {journal} {Phys. Rev. A}\ }\textbf {\bibinfo {volume} {82}},\
  \bibinfo {pages} {063419} (\bibinfo {year} {2010})}\BibitemShut {NoStop}%
\bibitem [{\citenamefont {Lange}\ \emph {et~al.}(2020)\citenamefont {Lange},
  \citenamefont {Huntemann}, \citenamefont {Sanner}, \citenamefont {Shao},
  \citenamefont {Lipphardt}, \citenamefont {Tamm},\ and\ \citenamefont
  {Peik}}]{lan20}%
  \BibitemOpen
  \bibfield  {author} {\bibinfo {author} {\bibfnamefont {R.}~\bibnamefont
  {Lange}}, \bibinfo {author} {\bibfnamefont {N.}~\bibnamefont {Huntemann}},
  \bibinfo {author} {\bibfnamefont {C.}~\bibnamefont {Sanner}}, \bibinfo
  {author} {\bibfnamefont {H.}~\bibnamefont {Shao}}, \bibinfo {author}
  {\bibfnamefont {B.}~\bibnamefont {Lipphardt}}, \bibinfo {author}
  {\bibfnamefont {C.}~\bibnamefont {Tamm}},\ and\ \bibinfo {author}
  {\bibfnamefont {E.}~\bibnamefont {Peik}},\ }\href
  {https://doi.org/10.1103/PhysRevLett.125.143201} {\bibfield  {journal}
  {\bibinfo  {journal} {Phys. Rev. Lett.}\ }\textbf {\bibinfo {volume} {125}},\
  \bibinfo {pages} {143201} (\bibinfo {year} {2020})}\BibitemShut {NoStop}%
\bibitem [{\citenamefont {Itano}(2000)}]{ita00}%
  \BibitemOpen
  \bibfield  {author} {\bibinfo {author} {\bibfnamefont {W.~M.}\ \bibnamefont
  {Itano}},\ }\href {https://doi.org/10.6028/jres.105.065} {\bibfield
  {journal} {\bibinfo  {journal} {J. Res. Natl. Inst. Stand. Technol.}\
  }\textbf {\bibinfo {volume} {105}},\ \bibinfo {pages} {829} (\bibinfo {year}
  {2000})},\ \bibinfo {note} {also see erratum: J. Res. NIST {\bf{11}}, 255
  (2006)}\BibitemShut {NoStop}%
\bibitem [{\citenamefont {Gan}\ \emph {et~al.}(2018)\citenamefont {Gan},
  \citenamefont {Maslennikov}, \citenamefont {Tseng}, \citenamefont {Tan},
  \citenamefont {Kaewuam}, \citenamefont {Arnold}, \citenamefont
  {Matsukevich},\ and\ \citenamefont {Barrett}}]{gan18}%
  \BibitemOpen
  \bibfield  {author} {\bibinfo {author} {\bibfnamefont {H.~C.~J.}\
  \bibnamefont {Gan}}, \bibinfo {author} {\bibfnamefont {G.}~\bibnamefont
  {Maslennikov}}, \bibinfo {author} {\bibfnamefont {K.-W.}\ \bibnamefont
  {Tseng}}, \bibinfo {author} {\bibfnamefont {T.~R.}\ \bibnamefont {Tan}},
  \bibinfo {author} {\bibfnamefont {R.}~\bibnamefont {Kaewuam}}, \bibinfo
  {author} {\bibfnamefont {K.~J.}\ \bibnamefont {Arnold}}, \bibinfo {author}
  {\bibfnamefont {D.}~\bibnamefont {Matsukevich}},\ and\ \bibinfo {author}
  {\bibfnamefont {M.~D.}\ \bibnamefont {Barrett}},\ }\href
  {https://doi.org/10.1103/PhysRevA.98.032514} {\bibfield  {journal} {\bibinfo
  {journal} {Phys. Rev. A}\ }\textbf {\bibinfo {volume} {98}},\ \bibinfo
  {pages} {032514} (\bibinfo {year} {2018})}\BibitemShut {NoStop}%
\bibitem [{\citenamefont {Baynham}\ \emph
  {et~al.}(2018{\natexlab{a}})\citenamefont {Baynham}, \citenamefont {Curtis},
  \citenamefont {Godun}, \citenamefont {Jones}, \citenamefont {Nisbet-Jones},
  \citenamefont {Baird}, \citenamefont {Bongs}, \citenamefont {Gill},
  \citenamefont {Fordell}, \citenamefont {Hieta}, \citenamefont {Lindvall},
  \citenamefont {Spidell},\ and\ \citenamefont {Lehman}}]{bay18}%
  \BibitemOpen
  \bibfield  {author} {\bibinfo {author} {\bibfnamefont {C.~F.~A.}\
  \bibnamefont {Baynham}}, \bibinfo {author} {\bibfnamefont {E.~A.}\
  \bibnamefont {Curtis}}, \bibinfo {author} {\bibfnamefont {R.~M.}\
  \bibnamefont {Godun}}, \bibinfo {author} {\bibfnamefont {J.~M.}\ \bibnamefont
  {Jones}}, \bibinfo {author} {\bibfnamefont {P.~B.~R.}\ \bibnamefont
  {Nisbet-Jones}}, \bibinfo {author} {\bibfnamefont {P.~E.~G.}\ \bibnamefont
  {Baird}}, \bibinfo {author} {\bibfnamefont {K.}~\bibnamefont {Bongs}},
  \bibinfo {author} {\bibfnamefont {P.}~\bibnamefont {Gill}}, \bibinfo {author}
  {\bibfnamefont {T.}~\bibnamefont {Fordell}}, \bibinfo {author} {\bibfnamefont
  {T.}~\bibnamefont {Hieta}}, \bibinfo {author} {\bibfnamefont
  {T.}~\bibnamefont {Lindvall}}, \bibinfo {author} {\bibfnamefont {M.~T.}\
  \bibnamefont {Spidell}},\ and\ \bibinfo {author} {\bibfnamefont {J.~H.}\
  \bibnamefont {Lehman}},\ }\href@noop {} {\bibinfo {title} {Measurement of
  differential polarizabilities at a mid-infrared wavelength in
  {$^{171}$Yb$^+$}}},\ \bibinfo {howpublished} {arXiv:1801.10134
  [physics.atom-ph]} (\bibinfo {year} {2018}{\natexlab{a}})\BibitemShut
  {NoStop}%
\bibitem [{\citenamefont {Tamm}\ \emph {et~al.}(2014)\citenamefont {Tamm},
  \citenamefont {Huntemann}, \citenamefont {Lipphardt}, \citenamefont
  {Gerginov}, \citenamefont {Nemitz}, \citenamefont {Kazda}, \citenamefont
  {Weyers},\ and\ \citenamefont {Peik}}]{tam14}%
  \BibitemOpen
  \bibfield  {author} {\bibinfo {author} {\bibfnamefont {C.}~\bibnamefont
  {Tamm}}, \bibinfo {author} {\bibfnamefont {N.}~\bibnamefont {Huntemann}},
  \bibinfo {author} {\bibfnamefont {B.}~\bibnamefont {Lipphardt}}, \bibinfo
  {author} {\bibfnamefont {V.}~\bibnamefont {Gerginov}}, \bibinfo {author}
  {\bibfnamefont {N.}~\bibnamefont {Nemitz}}, \bibinfo {author} {\bibfnamefont
  {M.}~\bibnamefont {Kazda}}, \bibinfo {author} {\bibfnamefont
  {S.}~\bibnamefont {Weyers}},\ and\ \bibinfo {author} {\bibfnamefont
  {E.}~\bibnamefont {Peik}},\ }\href
  {https://doi.org/10.1103/PhysRevA.89.023820} {\bibfield  {journal} {\bibinfo
  {journal} {Phys. Rev. A}\ }\textbf {\bibinfo {volume} {89}},\ \bibinfo
  {pages} {023820} (\bibinfo {year} {2014})}\BibitemShut {NoStop}%
\bibitem [{\citenamefont {Stenger}\ \emph {et~al.}(2002)\citenamefont
  {Stenger}, \citenamefont {Schnatz}, \citenamefont {Tamm},\ and\ \citenamefont
  {Telle}}]{ste02a}%
  \BibitemOpen
  \bibfield  {author} {\bibinfo {author} {\bibfnamefont {J.}~\bibnamefont
  {Stenger}}, \bibinfo {author} {\bibfnamefont {H.}~\bibnamefont {Schnatz}},
  \bibinfo {author} {\bibfnamefont {C.}~\bibnamefont {Tamm}},\ and\ \bibinfo
  {author} {\bibfnamefont {H.~R.}\ \bibnamefont {Telle}},\ }\href
  {https://doi.org/10.1103/PhysRevLett.88.073601} {\bibfield  {journal}
  {\bibinfo  {journal} {Phys. Rev. Lett.}\ }\textbf {\bibinfo {volume} {88}},\
  \bibinfo {pages} {073601} (\bibinfo {year} {2002})}\BibitemShut {NoStop}%
\bibitem [{\citenamefont {Huntemann}\ \emph {et~al.}(2016)\citenamefont
  {Huntemann}, \citenamefont {Sanner}, \citenamefont {Lipphardt}, \citenamefont
  {Tamm},\ and\ \citenamefont {Peik}}]{hun16}%
  \BibitemOpen
  \bibfield  {author} {\bibinfo {author} {\bibfnamefont {N.}~\bibnamefont
  {Huntemann}}, \bibinfo {author} {\bibfnamefont {C.}~\bibnamefont {Sanner}},
  \bibinfo {author} {\bibfnamefont {B.}~\bibnamefont {Lipphardt}}, \bibinfo
  {author} {\bibfnamefont {C.}~\bibnamefont {Tamm}},\ and\ \bibinfo {author}
  {\bibfnamefont {E.}~\bibnamefont {Peik}},\ }\href
  {https://doi.org/10.1103/PhysRevLett.116.063001} {\bibfield  {journal}
  {\bibinfo  {journal} {Phys. Rev. Lett.}\ }\textbf {\bibinfo {volume} {116}},\
  \bibinfo {pages} {063001} (\bibinfo {year} {2016})}\BibitemShut {NoStop}%
\bibitem [{\citenamefont {Weyers}\ \emph {et~al.}(2018)\citenamefont {Weyers},
  \citenamefont {Gerginov}, \citenamefont {Kazda}, \citenamefont {Rahm},
  \citenamefont {Lipphardt}, \citenamefont {Dobrev},\ and\ \citenamefont
  {Gibble}}]{wey18}%
  \BibitemOpen
  \bibfield  {author} {\bibinfo {author} {\bibfnamefont {S.}~\bibnamefont
  {Weyers}}, \bibinfo {author} {\bibfnamefont {V.}~\bibnamefont {Gerginov}},
  \bibinfo {author} {\bibfnamefont {M.}~\bibnamefont {Kazda}}, \bibinfo
  {author} {\bibfnamefont {J.}~\bibnamefont {Rahm}}, \bibinfo {author}
  {\bibfnamefont {B.}~\bibnamefont {Lipphardt}}, \bibinfo {author}
  {\bibfnamefont {G.}~\bibnamefont {Dobrev}},\ and\ \bibinfo {author}
  {\bibfnamefont {K.}~\bibnamefont {Gibble}},\ }\href
  {https://doi.org/10.1088/1681-7575/aae008} {\bibfield  {journal} {\bibinfo
  {journal} {Metrologia}\ }\textbf {\bibinfo {volume} {55}},\ \bibinfo {pages}
  {789} (\bibinfo {year} {2018})}\BibitemShut {NoStop}%
\bibitem [{\citenamefont {Leute}\ \emph {et~al.}(2016)\citenamefont {Leute},
  \citenamefont {Huntemann}, \citenamefont {Lipphardt}, \citenamefont {Tamm},
  \citenamefont {Nisbet-Jones}, \citenamefont {King}, \citenamefont {Godun},
  \citenamefont {Jones}, \citenamefont {Margolis}, \citenamefont {Whibberley},
  \citenamefont {Wallin}, \citenamefont {Merimaa}, \citenamefont {Gill},\ and\
  \citenamefont {Peik}}]{leu16}%
  \BibitemOpen
  \bibfield  {author} {\bibinfo {author} {\bibfnamefont {J.}~\bibnamefont
  {Leute}}, \bibinfo {author} {\bibfnamefont {N.}~\bibnamefont {Huntemann}},
  \bibinfo {author} {\bibfnamefont {B.}~\bibnamefont {Lipphardt}}, \bibinfo
  {author} {\bibfnamefont {C.}~\bibnamefont {Tamm}}, \bibinfo {author}
  {\bibfnamefont {P.~B.~R.}\ \bibnamefont {Nisbet-Jones}}, \bibinfo {author}
  {\bibfnamefont {S.~A.}\ \bibnamefont {King}}, \bibinfo {author}
  {\bibfnamefont {R.~M.}\ \bibnamefont {Godun}}, \bibinfo {author}
  {\bibfnamefont {J.~M.}\ \bibnamefont {Jones}}, \bibinfo {author}
  {\bibfnamefont {H.~S.}\ \bibnamefont {Margolis}}, \bibinfo {author}
  {\bibfnamefont {P.~B.}\ \bibnamefont {Whibberley}}, \bibinfo {author}
  {\bibfnamefont {A.}~\bibnamefont {Wallin}}, \bibinfo {author} {\bibfnamefont
  {M.}~\bibnamefont {Merimaa}}, \bibinfo {author} {\bibfnamefont
  {P.}~\bibnamefont {Gill}},\ and\ \bibinfo {author} {\bibfnamefont
  {E.}~\bibnamefont {Peik}},\ }\href
  {https://doi.org/10.1109/TUFFC.2016.2524988} {\bibfield  {journal} {\bibinfo
  {journal} {IEEE Trans. Ultrason. Ferroelectr. Freq. Control}\ }\textbf
  {\bibinfo {volume} {63}},\ \bibinfo {pages} {981} (\bibinfo {year}
  {2016})}\BibitemShut {NoStop}%
\bibitem [{\citenamefont {Grebing}\ \emph {et~al.}(2016)\citenamefont
  {Grebing}, \citenamefont {Al-Masoudi}, \citenamefont {D{\"o}rscher},
  \citenamefont {H{\"a}fner}, \citenamefont {Gerginov}, \citenamefont {Weyers},
  \citenamefont {Lipphardt}, \citenamefont {Riehle}, \citenamefont {Sterr},\
  and\ \citenamefont {Lisdat}}]{gre16}%
  \BibitemOpen
  \bibfield  {author} {\bibinfo {author} {\bibfnamefont {C.}~\bibnamefont
  {Grebing}}, \bibinfo {author} {\bibfnamefont {A.}~\bibnamefont {Al-Masoudi}},
  \bibinfo {author} {\bibfnamefont {S.}~\bibnamefont {D{\"o}rscher}}, \bibinfo
  {author} {\bibfnamefont {S.}~\bibnamefont {H{\"a}fner}}, \bibinfo {author}
  {\bibfnamefont {V.}~\bibnamefont {Gerginov}}, \bibinfo {author}
  {\bibfnamefont {S.}~\bibnamefont {Weyers}}, \bibinfo {author} {\bibfnamefont
  {B.}~\bibnamefont {Lipphardt}}, \bibinfo {author} {\bibfnamefont
  {F.}~\bibnamefont {Riehle}}, \bibinfo {author} {\bibfnamefont
  {U.}~\bibnamefont {Sterr}},\ and\ \bibinfo {author} {\bibfnamefont
  {C.}~\bibnamefont {Lisdat}},\ }\href
  {https://doi.org/10.1364/OPTICA.3.000563} {\bibfield  {journal} {\bibinfo
  {journal} {Optica}\ }\textbf {\bibinfo {volume} {3}},\ \bibinfo {pages} {563}
  (\bibinfo {year} {2016})}\BibitemShut {NoStop}%
\bibitem [{\citenamefont {Huntemann}\ \emph {et~al.}(2012)\citenamefont
  {Huntemann}, \citenamefont {Okhapkin}, \citenamefont {Lipphardt},
  \citenamefont {Weyers}, \citenamefont {Tamm},\ and\ \citenamefont
  {Peik}}]{hun12}%
  \BibitemOpen
  \bibfield  {author} {\bibinfo {author} {\bibfnamefont {N.}~\bibnamefont
  {Huntemann}}, \bibinfo {author} {\bibfnamefont {M.}~\bibnamefont {Okhapkin}},
  \bibinfo {author} {\bibfnamefont {B.}~\bibnamefont {Lipphardt}}, \bibinfo
  {author} {\bibfnamefont {S.}~\bibnamefont {Weyers}}, \bibinfo {author}
  {\bibfnamefont {C.}~\bibnamefont {Tamm}},\ and\ \bibinfo {author}
  {\bibfnamefont {E.}~\bibnamefont {Peik}},\ }\href
  {https://doi.org/10.1103/PhysRevLett.108.090801} {\bibfield  {journal}
  {\bibinfo  {journal} {Phys. Rev. Lett.}\ }\textbf {\bibinfo {volume} {108}},\
  \bibinfo {pages} {090801} (\bibinfo {year} {2012})}\BibitemShut {NoStop}%
\bibitem [{\citenamefont {Riedel}\ \emph {et~al.}(2020)\citenamefont {Riedel},
  \citenamefont {Al-Masoudi}, \citenamefont {Benkler}, \citenamefont
  {D\"orscher}, \citenamefont {Gerginov}, \citenamefont {Grebing},
  \citenamefont {H\"afner}, \citenamefont {Huntemann}, \citenamefont
  {Lipphardt}, \citenamefont {Lisdat}, \citenamefont {Peik}, \citenamefont
  {Piester}, \citenamefont {Sanner}, \citenamefont {Tamm}, \citenamefont
  {Weyers}, \citenamefont {Denker}, \citenamefont {Timmen}, \citenamefont
  {Voigt}, \citenamefont {Calonico}, \citenamefont {Cerretto}, \citenamefont
  {Costanzo}, \citenamefont {Levi}, \citenamefont {Sesia}, \citenamefont
  {Achkar}, \citenamefont {Gu\'ena}, \citenamefont {Abgrall}, \citenamefont
  {Rovera}, \citenamefont {Chupin}, \citenamefont {Shi}, \citenamefont
  {Bilicki}, \citenamefont {Bookjans}, \citenamefont {Lodewyck}, \citenamefont
  {Le~Targat}, \citenamefont {Delva}, \citenamefont {Bize}, \citenamefont
  {Baynes}, \citenamefont {Baynham}, \citenamefont {Bowden}, \citenamefont
  {Gill}, \citenamefont {Godun}, \citenamefont {Hill}, \citenamefont {Hobson},
  \citenamefont {Jones}, \citenamefont {King}, \citenamefont {Nisbet-Jones},
  \citenamefont {Rolland}, \citenamefont {Shemar}, \citenamefont {Whibberley},\
  and\ \citenamefont {Margolis}}]{rie20}%
  \BibitemOpen
  \bibfield  {author} {\bibinfo {author} {\bibfnamefont {F.}~\bibnamefont
  {Riedel}}, \bibinfo {author} {\bibfnamefont {A.}~\bibnamefont {Al-Masoudi}},
  \bibinfo {author} {\bibfnamefont {E.}~\bibnamefont {Benkler}}, \bibinfo
  {author} {\bibfnamefont {S.}~\bibnamefont {D\"orscher}}, \bibinfo {author}
  {\bibfnamefont {V.}~\bibnamefont {Gerginov}}, \bibinfo {author}
  {\bibfnamefont {C.}~\bibnamefont {Grebing}}, \bibinfo {author} {\bibfnamefont
  {S.}~\bibnamefont {H\"afner}}, \bibinfo {author} {\bibfnamefont
  {N.}~\bibnamefont {Huntemann}}, \bibinfo {author} {\bibfnamefont
  {B.}~\bibnamefont {Lipphardt}}, \bibinfo {author} {\bibfnamefont
  {C.}~\bibnamefont {Lisdat}}, \bibinfo {author} {\bibfnamefont
  {E.}~\bibnamefont {Peik}}, \bibinfo {author} {\bibfnamefont {D.}~\bibnamefont
  {Piester}}, \bibinfo {author} {\bibfnamefont {C.}~\bibnamefont {Sanner}},
  \bibinfo {author} {\bibfnamefont {C.}~\bibnamefont {Tamm}}, \bibinfo {author}
  {\bibfnamefont {S.}~\bibnamefont {Weyers}}, \bibinfo {author} {\bibfnamefont
  {H.}~\bibnamefont {Denker}}, \bibinfo {author} {\bibfnamefont
  {L.}~\bibnamefont {Timmen}}, \bibinfo {author} {\bibfnamefont
  {C.}~\bibnamefont {Voigt}}, \bibinfo {author} {\bibfnamefont
  {D.}~\bibnamefont {Calonico}}, \bibinfo {author} {\bibfnamefont
  {G.}~\bibnamefont {Cerretto}}, \bibinfo {author} {\bibfnamefont {G.~A.}\
  \bibnamefont {Costanzo}}, \bibinfo {author} {\bibfnamefont {F.}~\bibnamefont
  {Levi}}, \bibinfo {author} {\bibfnamefont {I.}~\bibnamefont {Sesia}},
  \bibinfo {author} {\bibfnamefont {J.}~\bibnamefont {Achkar}}, \bibinfo
  {author} {\bibfnamefont {J.}~\bibnamefont {Gu\'ena}}, \bibinfo {author}
  {\bibfnamefont {M.}~\bibnamefont {Abgrall}}, \bibinfo {author} {\bibfnamefont
  {G.~D.}\ \bibnamefont {Rovera}}, \bibinfo {author} {\bibfnamefont
  {B.}~\bibnamefont {Chupin}}, \bibinfo {author} {\bibfnamefont
  {C.}~\bibnamefont {Shi}}, \bibinfo {author} {\bibfnamefont {S.}~\bibnamefont
  {Bilicki}}, \bibinfo {author} {\bibfnamefont {E.}~\bibnamefont {Bookjans}},
  \bibinfo {author} {\bibfnamefont {J.}~\bibnamefont {Lodewyck}}, \bibinfo
  {author} {\bibfnamefont {R.}~\bibnamefont {Le~Targat}}, \bibinfo {author}
  {\bibfnamefont {P.}~\bibnamefont {Delva}}, \bibinfo {author} {\bibfnamefont
  {S.}~\bibnamefont {Bize}}, \bibinfo {author} {\bibfnamefont {F.~N.}\
  \bibnamefont {Baynes}}, \bibinfo {author} {\bibfnamefont {C.}~\bibnamefont
  {Baynham}}, \bibinfo {author} {\bibfnamefont {W.}~\bibnamefont {Bowden}},
  \bibinfo {author} {\bibfnamefont {P.}~\bibnamefont {Gill}}, \bibinfo {author}
  {\bibfnamefont {R.~M.}\ \bibnamefont {Godun}}, \bibinfo {author}
  {\bibfnamefont {I.~R.}\ \bibnamefont {Hill}}, \bibinfo {author}
  {\bibfnamefont {R.}~\bibnamefont {Hobson}}, \bibinfo {author} {\bibfnamefont
  {J.~M.}\ \bibnamefont {Jones}}, \bibinfo {author} {\bibfnamefont {S.~A.}\
  \bibnamefont {King}}, \bibinfo {author} {\bibfnamefont {P.}~\bibnamefont
  {Nisbet-Jones}}, \bibinfo {author} {\bibfnamefont {A.}~\bibnamefont
  {Rolland}}, \bibinfo {author} {\bibfnamefont {S.~L.}\ \bibnamefont {Shemar}},
  \bibinfo {author} {\bibfnamefont {P.~B.}\ \bibnamefont {Whibberley}},\ and\
  \bibinfo {author} {\bibfnamefont {H.~S.}\ \bibnamefont {Margolis}},\ }\href
  {https://doi.org/10.1088/1681-7575/ab6745} {\bibfield  {journal} {\bibinfo
  {journal} {Metrologia}\ }\textbf {\bibinfo {volume} {57}},\ \bibinfo {pages}
  {045005} (\bibinfo {year} {2020})}\BibitemShut {NoStop}%
\bibitem [{\citenamefont {Riehle}\ \emph {et~al.}(2018)\citenamefont {Riehle},
  \citenamefont {Gill}, \citenamefont {Arias},\ and\ \citenamefont
  {Robertsson}}]{rie18}%
  \BibitemOpen
  \bibfield  {author} {\bibinfo {author} {\bibfnamefont {F.}~\bibnamefont
  {Riehle}}, \bibinfo {author} {\bibfnamefont {P.}~\bibnamefont {Gill}},
  \bibinfo {author} {\bibfnamefont {F.}~\bibnamefont {Arias}},\ and\ \bibinfo
  {author} {\bibfnamefont {L.}~\bibnamefont {Robertsson}},\ }\href
  {https://doi.org/10.1088/1681-7575/aaa302} {\bibfield  {journal} {\bibinfo
  {journal} {Metrologia}\ }\textbf {\bibinfo {volume} {55}},\ \bibinfo {pages}
  {188} (\bibinfo {year} {2018})}\BibitemShut {NoStop}%
\bibitem [{\citenamefont {Schwarz}\ \emph {et~al.}(2020)\citenamefont
  {Schwarz}, \citenamefont {D\"{o}rscher}, \citenamefont {Al-Masoudi},
  \citenamefont {Benkler}, \citenamefont {Legero}, \citenamefont {Sterr},
  \citenamefont {Weyers}, \citenamefont {Rahm}, \citenamefont {Lipphardt},\
  and\ \citenamefont {Lisdat}}]{sch20d}%
  \BibitemOpen
  \bibfield  {author} {\bibinfo {author} {\bibfnamefont {R.}~\bibnamefont
  {Schwarz}}, \bibinfo {author} {\bibfnamefont {S.}~\bibnamefont
  {D\"{o}rscher}}, \bibinfo {author} {\bibfnamefont {A.}~\bibnamefont
  {Al-Masoudi}}, \bibinfo {author} {\bibfnamefont {E.}~\bibnamefont {Benkler}},
  \bibinfo {author} {\bibfnamefont {T.}~\bibnamefont {Legero}}, \bibinfo
  {author} {\bibfnamefont {U.}~\bibnamefont {Sterr}}, \bibinfo {author}
  {\bibfnamefont {S.}~\bibnamefont {Weyers}}, \bibinfo {author} {\bibfnamefont
  {J.}~\bibnamefont {Rahm}}, \bibinfo {author} {\bibfnamefont {B.}~\bibnamefont
  {Lipphardt}},\ and\ \bibinfo {author} {\bibfnamefont {C.}~\bibnamefont
  {Lisdat}},\ }\href {https://doi.org/10.1103/PhysRevResearch.2.033242}
  {\bibfield  {journal} {\bibinfo  {journal} {Phys. Rev. Research}\ }\textbf
  {\bibinfo {volume} {2}},\ \bibinfo {pages} {033242} (\bibinfo {year}
  {2020})}\BibitemShut {NoStop}%
\bibitem [{\citenamefont {Baynham}\ \emph
  {et~al.}(2018{\natexlab{b}})\citenamefont {Baynham}, \citenamefont {Godun},
  \citenamefont {Jones}, \citenamefont {King}, \citenamefont {Nisbet-Jones},
  \citenamefont {Baynes}, \citenamefont {Rolland}, \citenamefont {Baird},
  \citenamefont {Bongs}, \citenamefont {Gill},\ and\ \citenamefont
  {Margolis}}]{bay18a}%
  \BibitemOpen
  \bibfield  {author} {\bibinfo {author} {\bibfnamefont {C.~F.~A.}\
  \bibnamefont {Baynham}}, \bibinfo {author} {\bibfnamefont {R.~M.}\
  \bibnamefont {Godun}}, \bibinfo {author} {\bibfnamefont {J.~M.}\ \bibnamefont
  {Jones}}, \bibinfo {author} {\bibfnamefont {S.~A.}\ \bibnamefont {King}},
  \bibinfo {author} {\bibfnamefont {P.~B.~R.}\ \bibnamefont {Nisbet-Jones}},
  \bibinfo {author} {\bibfnamefont {F.}~\bibnamefont {Baynes}}, \bibinfo
  {author} {\bibfnamefont {A.}~\bibnamefont {Rolland}}, \bibinfo {author}
  {\bibfnamefont {P.~E.~G.}\ \bibnamefont {Baird}}, \bibinfo {author}
  {\bibfnamefont {K.}~\bibnamefont {Bongs}}, \bibinfo {author} {\bibfnamefont
  {P.}~\bibnamefont {Gill}},\ and\ \bibinfo {author} {\bibfnamefont {H.~S.}\
  \bibnamefont {Margolis}},\ }\href
  {https://doi.org/10.1080/09500340.2017.1384514} {\bibfield  {journal}
  {\bibinfo  {journal} {J. Mod. Optics}\ }\textbf {\bibinfo {volume} {65}},\
  \bibinfo {pages} {585} (\bibinfo {year} {2018}{\natexlab{b}})}\BibitemShut
  {NoStop}%
\bibitem [{\citenamefont {McGrew}\ \emph {et~al.}(2019)\citenamefont {McGrew},
  \citenamefont {Zhang}, \citenamefont {Leopardi}, \citenamefont {Fasano},
  \citenamefont {Nicolodi}, \citenamefont {Beloy}, \citenamefont {Yao},
  \citenamefont {Sherman}, \citenamefont {Sch\"{a}ffer}, \citenamefont
  {Savory}, \citenamefont {Brown}, \citenamefont {R\"{o}misch}, \citenamefont
  {Oates}, \citenamefont {Parker}, \citenamefont {Fortier},\ and\ \citenamefont
  {Ludlow}}]{mcg19}%
  \BibitemOpen
  \bibfield  {author} {\bibinfo {author} {\bibfnamefont {W.~F.}\ \bibnamefont
  {McGrew}}, \bibinfo {author} {\bibfnamefont {X.}~\bibnamefont {Zhang}},
  \bibinfo {author} {\bibfnamefont {H.}~\bibnamefont {Leopardi}}, \bibinfo
  {author} {\bibfnamefont {R.~J.}\ \bibnamefont {Fasano}}, \bibinfo {author}
  {\bibfnamefont {D.}~\bibnamefont {Nicolodi}}, \bibinfo {author}
  {\bibfnamefont {K.}~\bibnamefont {Beloy}}, \bibinfo {author} {\bibfnamefont
  {J.}~\bibnamefont {Yao}}, \bibinfo {author} {\bibfnamefont {J.~A.}\
  \bibnamefont {Sherman}}, \bibinfo {author} {\bibfnamefont {S.~A.}\
  \bibnamefont {Sch\"{a}ffer}}, \bibinfo {author} {\bibfnamefont
  {J.}~\bibnamefont {Savory}}, \bibinfo {author} {\bibfnamefont {R.~C.}\
  \bibnamefont {Brown}}, \bibinfo {author} {\bibfnamefont {S.}~\bibnamefont
  {R\"{o}misch}}, \bibinfo {author} {\bibfnamefont {C.~W.}\ \bibnamefont
  {Oates}}, \bibinfo {author} {\bibfnamefont {T.~E.}\ \bibnamefont {Parker}},
  \bibinfo {author} {\bibfnamefont {T.~M.}\ \bibnamefont {Fortier}},\ and\
  \bibinfo {author} {\bibfnamefont {A.~D.}\ \bibnamefont {Ludlow}},\ }\href
  {https://doi.org/10.1364/OPTICA.6.000448} {\bibfield  {journal} {\bibinfo
  {journal} {Optica}\ }\textbf {\bibinfo {volume} {6}},\ \bibinfo {pages} {448}
  (\bibinfo {year} {2019})}\BibitemShut {NoStop}%
\bibitem [{\citenamefont {Nemitz}\ \emph {et~al.}(2020)\citenamefont {Nemitz},
  \citenamefont {Gotoh}, \citenamefont {Nakagawa}, \citenamefont {Ito},
  \citenamefont {Hanado}, \citenamefont {Ido},\ and\ \citenamefont
  {Hachisu}}]{nem20}%
  \BibitemOpen
  \bibfield  {author} {\bibinfo {author} {\bibfnamefont {N.}~\bibnamefont
  {Nemitz}}, \bibinfo {author} {\bibfnamefont {T.}~\bibnamefont {Gotoh}},
  \bibinfo {author} {\bibfnamefont {F.}~\bibnamefont {Nakagawa}}, \bibinfo
  {author} {\bibfnamefont {H.}~\bibnamefont {Ito}}, \bibinfo {author}
  {\bibfnamefont {Y.}~\bibnamefont {Hanado}}, \bibinfo {author} {\bibfnamefont
  {T.}~\bibnamefont {Ido}},\ and\ \bibinfo {author} {\bibfnamefont
  {H.}~\bibnamefont {Hachisu}},\ }\bibfield  {journal} {\bibinfo  {journal}
  {Metrologia}\ }\href {https://doi.org/10.1088/1681-7575/abc232}
  {10.1088/1681-7575/abc232} (\bibinfo {year} {2020})\BibitemShut {NoStop}%
\bibitem [{\citenamefont {Dinh}\ \emph {et~al.}(2009)\citenamefont {Dinh},
  \citenamefont {Dunning}, \citenamefont {Dzuba},\ and\ \citenamefont
  {Flambaum}}]{din09}%
  \BibitemOpen
  \bibfield  {author} {\bibinfo {author} {\bibfnamefont {T.~H.}\ \bibnamefont
  {Dinh}}, \bibinfo {author} {\bibfnamefont {A.}~\bibnamefont {Dunning}},
  \bibinfo {author} {\bibfnamefont {V.~A.}\ \bibnamefont {Dzuba}},\ and\
  \bibinfo {author} {\bibfnamefont {V.~V.}\ \bibnamefont {Flambaum}},\ }\href
  {https://doi.org/10.1103/PhysRevA.79.054102} {\bibfield  {journal} {\bibinfo
  {journal} {Phys. Rev. A}\ }\textbf {\bibinfo {volume} {79}},\ \bibinfo
  {pages} {054102} (\bibinfo {year} {2009})}\BibitemShut {NoStop}%
\bibitem [{\citenamefont {Dzuba}\ and\ \citenamefont {Flambaum}(2017)}]{dzu17}%
  \BibitemOpen
  \bibfield  {author} {\bibinfo {author} {\bibfnamefont {V.~A.}\ \bibnamefont
  {Dzuba}}\ and\ \bibinfo {author} {\bibfnamefont {V.~V.}\ \bibnamefont
  {Flambaum}},\ }\href {https://doi.org/10.1103/PhysRevD.95.015019} {\bibfield
  {journal} {\bibinfo  {journal} {Phys. Rev. D}\ }\textbf {\bibinfo {volume}
  {95}},\ \bibinfo {pages} {015019} (\bibinfo {year} {2017})}\BibitemShut
  {NoStop}%
\bibitem [{\citenamefont {Arvanitaki}\ \emph {et~al.}(2015)\citenamefont
  {Arvanitaki}, \citenamefont {Huang},\ and\ \citenamefont
  {Van~Tilburg}}]{arv15}%
  \BibitemOpen
  \bibfield  {author} {\bibinfo {author} {\bibfnamefont {A.}~\bibnamefont
  {Arvanitaki}}, \bibinfo {author} {\bibfnamefont {J.}~\bibnamefont {Huang}},\
  and\ \bibinfo {author} {\bibfnamefont {K.}~\bibnamefont {Van~Tilburg}},\
  }\href {https://doi.org/10.1103/PhysRevD.91.015015} {\bibfield  {journal}
  {\bibinfo  {journal} {Phys. Rev. D}\ }\textbf {\bibinfo {volume} {91}},\
  \bibinfo {pages} {015015} (\bibinfo {year} {2015})}\BibitemShut {NoStop}%
\bibitem [{\citenamefont {Roberts}\ \emph {et~al.}(2020)\citenamefont
  {Roberts}, \citenamefont {Delva}, \citenamefont {Al-Masoudi}, \citenamefont
  {Amy-Klein}, \citenamefont {B{\ae}rentsen}, \citenamefont {Baynham},
  \citenamefont {Benkler}, \citenamefont {Bilicki}, \citenamefont {Bowden},
  \citenamefont {Cantin}, \citenamefont {Curtis}, \citenamefont {D\"orscher},
  \citenamefont {Frank}, \citenamefont {Gill}, \citenamefont {Godun},
  \citenamefont {Grosche}, \citenamefont {Hees}, \citenamefont {Hill},
  \citenamefont {Hobson}, \citenamefont {Huntemann}, \citenamefont
  {Kronj\"ager}, \citenamefont {Koke}, \citenamefont {Kuhl}, \citenamefont
  {Lange}, \citenamefont {Legero}, \citenamefont {Lipphardt}, \citenamefont
  {Lisdat}, \citenamefont {Lodewyck}, \citenamefont {Lopez}, \citenamefont
  {Margolis}, \citenamefont {\'Alvarez-Mart\'inez}, \citenamefont {Meynadier},
  \citenamefont {Ozimek}, \citenamefont {Peik}, \citenamefont {Pottie},
  \citenamefont {Quintin}, \citenamefont {Schwarz}, \citenamefont {Sanner},
  \citenamefont {Schioppo}, \citenamefont {Silva}, \citenamefont {Sterr},
  \citenamefont {Tamm}, \citenamefont {LeTargat}, \citenamefont {Tuckey},
  \citenamefont {Vallet}, \citenamefont {Waterholter}, \citenamefont {Xu},\
  and\ \citenamefont {Wolf}}]{rob20}%
  \BibitemOpen
  \bibfield  {author} {\bibinfo {author} {\bibfnamefont {B.~M.}\ \bibnamefont
  {Roberts}}, \bibinfo {author} {\bibfnamefont {P.}~\bibnamefont {Delva}},
  \bibinfo {author} {\bibfnamefont {A.}~\bibnamefont {Al-Masoudi}}, \bibinfo
  {author} {\bibfnamefont {A.}~\bibnamefont {Amy-Klein}}, \bibinfo {author}
  {\bibfnamefont {C.}~\bibnamefont {B{\ae}rentsen}}, \bibinfo {author}
  {\bibfnamefont {C.~F.~A.}\ \bibnamefont {Baynham}}, \bibinfo {author}
  {\bibfnamefont {E.}~\bibnamefont {Benkler}}, \bibinfo {author} {\bibfnamefont
  {S.}~\bibnamefont {Bilicki}}, \bibinfo {author} {\bibfnamefont
  {W.}~\bibnamefont {Bowden}}, \bibinfo {author} {\bibfnamefont
  {E.}~\bibnamefont {Cantin}}, \bibinfo {author} {\bibfnamefont {E.~A.}\
  \bibnamefont {Curtis}}, \bibinfo {author} {\bibfnamefont {S.}~\bibnamefont
  {D\"orscher}}, \bibinfo {author} {\bibfnamefont {F.}~\bibnamefont {Frank}},
  \bibinfo {author} {\bibfnamefont {P.}~\bibnamefont {Gill}}, \bibinfo {author}
  {\bibfnamefont {R.~M.}\ \bibnamefont {Godun}}, \bibinfo {author}
  {\bibfnamefont {G.}~\bibnamefont {Grosche}}, \bibinfo {author} {\bibfnamefont
  {A.}~\bibnamefont {Hees}}, \bibinfo {author} {\bibfnamefont {I.~R.}\
  \bibnamefont {Hill}}, \bibinfo {author} {\bibfnamefont {R.}~\bibnamefont
  {Hobson}}, \bibinfo {author} {\bibfnamefont {N.}~\bibnamefont {Huntemann}},
  \bibinfo {author} {\bibfnamefont {J.}~\bibnamefont {Kronj\"ager}}, \bibinfo
  {author} {\bibfnamefont {S.}~\bibnamefont {Koke}}, \bibinfo {author}
  {\bibfnamefont {A.}~\bibnamefont {Kuhl}}, \bibinfo {author} {\bibfnamefont
  {R.}~\bibnamefont {Lange}}, \bibinfo {author} {\bibfnamefont
  {T.}~\bibnamefont {Legero}}, \bibinfo {author} {\bibfnamefont
  {B.}~\bibnamefont {Lipphardt}}, \bibinfo {author} {\bibfnamefont
  {C.}~\bibnamefont {Lisdat}}, \bibinfo {author} {\bibfnamefont
  {J.}~\bibnamefont {Lodewyck}}, \bibinfo {author} {\bibfnamefont
  {O.}~\bibnamefont {Lopez}}, \bibinfo {author} {\bibfnamefont {H.~S.}\
  \bibnamefont {Margolis}}, \bibinfo {author} {\bibfnamefont {H.}~\bibnamefont
  {\'Alvarez-Mart\'inez}}, \bibinfo {author} {\bibfnamefont {F.}~\bibnamefont
  {Meynadier}}, \bibinfo {author} {\bibfnamefont {F.}~\bibnamefont {Ozimek}},
  \bibinfo {author} {\bibfnamefont {E.}~\bibnamefont {Peik}}, \bibinfo {author}
  {\bibfnamefont {P.-E.}\ \bibnamefont {Pottie}}, \bibinfo {author}
  {\bibfnamefont {N.}~\bibnamefont {Quintin}}, \bibinfo {author} {\bibfnamefont
  {R.}~\bibnamefont {Schwarz}}, \bibinfo {author} {\bibfnamefont
  {C.}~\bibnamefont {Sanner}}, \bibinfo {author} {\bibfnamefont
  {M.}~\bibnamefont {Schioppo}}, \bibinfo {author} {\bibfnamefont
  {A.}~\bibnamefont {Silva}}, \bibinfo {author} {\bibfnamefont
  {U.}~\bibnamefont {Sterr}}, \bibinfo {author} {\bibfnamefont
  {C.}~\bibnamefont {Tamm}}, \bibinfo {author} {\bibfnamefont {R.}~\bibnamefont
  {LeTargat}}, \bibinfo {author} {\bibfnamefont {P.}~\bibnamefont {Tuckey}},
  \bibinfo {author} {\bibfnamefont {G.}~\bibnamefont {Vallet}}, \bibinfo
  {author} {\bibfnamefont {T.}~\bibnamefont {Waterholter}}, \bibinfo {author}
  {\bibfnamefont {D.}~\bibnamefont {Xu}},\ and\ \bibinfo {author}
  {\bibfnamefont {P.}~\bibnamefont {Wolf}},\ }\href
  {https://doi.org/10.1088/1367-2630/abaace} {\bibfield  {journal} {\bibinfo
  {journal} {New J. Phys.}\ }\textbf {\bibinfo {volume} {22}},\ \bibinfo
  {pages} {093010} (\bibinfo {year} {2020})}\BibitemShut {NoStop}%
\end{thebibliography}
\end{document}